\documentclass[12pt]{article}
\usepackage{amsmath}
\usepackage{amsfonts}
\usepackage{amssymb}
\usepackage{mathtools}
\usepackage{mathrsfs}
\usepackage{bbm}
\usepackage{dsfont}
\usepackage{bm}
\usepackage{graphicx}
\usepackage{subcaption}
\usepackage{booktabs}
\usepackage{pifont}
\usepackage{natbib}
\usepackage{url}
\usepackage[dvipsnames]{xcolor}
\usepackage{tikz}
\newcommand{\bfa}{\bm{a}}
\newcommand{\bfb}{\bm{b}}
\newcommand{\bfc}{\bm{c}}

\newcommand{\bfq}{\bm{q}}

\newcommand{\bft}{\bm{t}}
\newcommand{\bfu}{\bm{u}}

\newcommand{\bfbeta}{\bm{\beta}}
\newcommand{\bfgamma}{\bm{\gamma}}

\newcommand{\bftheta}{\bm{\theta}}

\newcommand{\bfDelta}{\bm{\Delta}}
\newcommand{\bfGamma}{\bm{\Gamma}}

\newcommand{\bfSigma}{\bm{\Sigma}}

\newcommand{\bfOmega}{\bm{\Omega}}

\newcommand{\rmC}{\mathrm{C}}
\newcommand{\rmD}{\mathrm{D}}
\newcommand{\rmR}{\mathrm{R}}
\newcommand{\rmU}{\mathrm{U}}

\newcommand{\rmd}{\mathrm{d}}

\newcommand{\bfI}{\bm{I}}

\newcommand{\bfT}{\bm{T}}
\newcommand{\bfU}{\bm{U}}
\newcommand{\bfV}{\bm{V}}

\newcommand{\bfZ}{\bm{Z}}

\newcommand{\bbE}{\mathbb{E}}

\newcommand{\bbN}{\mathbb{N}}

\newcommand{\bbP}{\mathbb{P}}
\newcommand{\bbR}{\mathbb{R}}

\newcommand{\bbone}{\mathbbm{1}}

\newcommand{\bbmc}{\mathbbm{c}}

\newcommand{\bbmC}{\mathbbm{C}}

\newcommand{\calE}{\mathcal{E}}

\newcommand{\calO}{\mathcal{O}}

\newcommand{\calT}{\mathcal{T}}
\newcommand{\calU}{\mathcal{U}}

\newcommand{\frako}{\mathfrak{o}}
\newcommand{\frakc}{\mathfrak{c}}

\newcommand{\frakn}{\mathfrak{n}}

\newcommand{\frakA}{\mathfrak{A}}

\newcommand{\frakD}{\mathfrak{D}}

\newcommand{\frakJ}{\mathfrak{J}}

\newcommand{\frakR}{\mathfrak{R}}

\newcommand{\bfthetahat}{\widehat{\bftheta}}

\newcommand{\Uhat}{\widehat{U}}
\newcommand{\bfUhat}{\widehat{\bfU}}
\newcommand{\calUhat}{\widehat{\calU}}
\newcommand{\bfbetahat}{\widehat{\bfbeta}}
\newcommand{\Lambdahat}{\widehat{\Lambda}}
\newcommand{\bfgammahat}{\widehat{\bfgamma}}
\newcommand{\bfGammahat}{\widehat{\bfGamma}}

\newcommand{\scrE}{\mathscr{E}}

\newcommand{\scrL}{\mathscr{L}}
\newcommand{\scrN}{\mathscr{N}}
\newcommand{\scrT}{\mathscr{T}}
\newcommand{\scrV}{\mathscr{V}}

\newtheorem{proposition}{Proposition}
\newtheorem{corollary}{Corollary}
\newtheorem{remark}{Remark}

\usepackage[colorlinks=true,linkcolor=blue,citecolor=black]{hyperref}

\newcommand{\blind}{0}

\allowdisplaybreaks

\begin{document}

\def\spacingset#1{\renewcommand{\baselinestretch}%
{#1}\small\normalsize} \spacingset{1}


\if0\blind
{
  \begin{center}
	{\Large \bf Stage-wise semiparametric analysis of multivariate event times under informative censoring via vine copula}\\[0.5cm]
	{\large Xinyuan Chen$^{1,\dagger}$ ~ Yiwei Li$^{2}$ ~ Qian M. Zhou$^{1,\dagger,\ast}$}\\[0.2cm]
	$^1$Department of Mathematics and Statistics, Mississippi State University, \\Mississippi State, MS, USA\\
	$^2$Department of Marketing and International Business, Lingnan University, \\Tuen Mun, New Territories, HKSAR, China\\
	$^\dagger$Xinyuan Chen and Qian M. Zhou contributed equally and share the first authorship.\\
	$^\ast$qz70@msstate.edu
  \end{center}
} \fi

\if1\blind
{
  \bigskip
  \bigskip
  \bigskip
  \begin{center}
    {\Large \bf Stage-wise semiparametric analysis of multivariate event times under informative censoring via vine copula}
\end{center}
  \medskip
} \fi

\begin{abstract}
    The analysis of times to multiple nonterminal events arises in many lifetime data settings, where the nonterminal event times are subject to informative censoring by a terminal event, posing substantial challenges. We propose a likelihood-based regression framework that models dependence among multiple nonterminal and terminal event times using a vine copula. This framework allows pairwise associations between nonterminal and terminal events, as well as among nonterminal events, to vary across event types, while enabling valid regression inference for the marginal distributions of individual event times. To facilitate computation, we develop a stage-wise pseudo-likelihood estimation procedure and establish its large-sample properties. We demonstrate satisfactory finite-sample performance of the proposed estimators and associated variance estimators through simulation studies. We also illustrate the proposed methodology using data from a crowdfunding platform to examine the relationship between different types of creator-backer interactions and a creator's lifetime on the platform.
\end{abstract}

\noindent%

\spacingset{1.5}

\section{Introduction} \label{sec:intro}

Multivariate event times under informative censoring arise when multiple nonterminal events are subject to censoring by a terminal event, with the censoring mechanism being informative due to possible dependence between nonterminal and terminal event times. For example, in business studies of crowdfunding platforms \citep{Dai2019,Su2023}, comments from campaign backers and updates from creators are nonterminal events of interest that may be censored when a creator leaves the platform. Similarly, in breast cancer survivorship studies \citep{Davis2014}, survivors may experience multiple nonterminal events, such as disease relapse, second primary cancers, or cardiovascular complications, which are subject to informative censoring by death, as well as administrative censoring at the end of follow-up. In analyzing such data, primary inferential goals often include characterizing pairwise associations between event times, both between nonterminal and terminal events and among nonterminal events, as well as estimating covariate effects on individual event times through regression models.

Existing literature on informative censoring predominantly evaluates semi-competing risks, wherein a single nonterminal event may be precluded by a terminal event \citep{Fine2001, peng2007regression, hsieh2008regression, Chen2012, zhu2021semiparametric, Arachchige2024}. A parallel framework exists for recurrent event processes terminated by a dependent competing risk. To characterize the sequential dependence between recurrent gap times and the terminal event, this framework includes shared frailty models \citep{Sun2024}, copula-based Markov chain models \citep{Huang2021}, and Bayesian hierarchical joint copula models for time-varying correlations and dynamic prediction \citep{Li2019Bayesian, Li2020Bayesian, Liang2023}. Furthermore, vine copulas have been applied in survival analysis to factorize complex dependencies, specifically for clustered event times subject to a common right-censoring \citep{Barthel2018}, unbalanced recurrent gap times under induced dependent censoring \citep{Barthel2019}, and single right-censored responses conditional on mixed discrete-continuous covariates \citep{Pan2022}.

Despite these developments, methodologies for jointly analyzing multiple distinct nonterminal events under informative censoring remain limited. Unlike recurrent events—where subjects experience repeated occurrences of the same event type—multivariate nonterminal events involve distinct event types without an inherent ordering. Consequently, models based on shared frailties or two-level copulas that are designed for sequential gap-time dependence cannot adequately capture the heterogeneous pairwise dependence structure among distinct multivariate events. For this setting, \citet{Li2020} used a single multivariate copula to specify the joint survival function of all event times, yielding homogeneous pairwise dependencies among them.  \citet{Li2023} utilized a two-level nested copula that distinguishes the dependence between nonterminal and terminal events from that among nonterminal events.

In this article, we present a complementary flexible analysis framework specifically designed for multivariate event times under informative censoring. The proposed framework leverages vine copulas \citep{Joe2014} to decompose the complex multivariate dependence among all event times into a collection of pairwise dependencies organized through a sequence of trees (connected acyclic graphs). Each pairwise dependence is represented by a bivariate copula, allowing different event pairs to be modeled using different copula families and dependence parameters. Consequently, the framework accommodates heterogeneous associations between nonterminal and terminal event times as well as between pairs of nonterminal event times.

Because the terminal event informatively censors every nonterminal event, this mechanism leads us to adopt a star-shaped first tree, with the root node representing the terminal event time and connected to all nonterminal event times (Figure~\ref{fig:vg-5}). This construction of the first tree, together with the proposed stage-wise estimation procedure, enables direct regression modeling of the nonterminal marginal survival distributions while accounting for their dependence with the terminal event, thereby yielding consistent estimation of the marginal model parameters. For subsequent trees, users are free to specify the vine structure according to the scientific context, allowing the remaining dependence structure to be modeled with considerable flexibility.

Simultaneous estimation of all model parameters is computationally challenging. Therefore, we develop a computationally efficient stage-wise pseudo maximum likelihood estimation (PMLE) procedure to estimate the parameters associated with each tree sequentially. We establish the asymptotic properties of the resulting estimators and provide analytical variance estimators, the validity of which is demonstrated through Simulation Study I. It is worth noting that the first-stage estimation is modular, with each marginal distribution and its corresponding first-tree copula parameter estimated separately through a pairwise likelihood. This separation of estimation limits the impact of misspecifying one first-tree copula family on the estimation of the remaining marginal distributions and first-tree copula parameters. This localized robustness is demonstrated in Simulation Study II.

The remainder of the article is organized as follows. Section~\ref{sec:mod} formulates the data structure and presents the vine copula-based method. Section~\ref{sec:est} gives the stage-wise estimation and inference procedure. Section~\ref{sec:sim} covers results from simulation studies. Section~\ref{sec:data} illustrates the application of the proposed methodology to a real data example from the crowdfunding platform. Section~\ref{sec:disc} concludes with a discussion. Proofs of all theoretical results are available in the Supplementary Materials (Supp.). The proposed method is implemented in the \texttt{R} package \texttt{VineSurvIC} (\url{https://github.com/michellezhou2009/VineSurvIC}). \texttt{R} code and documentation for reproducing the simulation studies are available in the GitHub repository \url{https://github.com/michellezhou2009/VineMETIC}.

Regarding notations, $\bfT_{\frakJ}=(T_j, j\in \frakJ)^\top$ denotes a vector of variables for a nonempty subset $\frakJ\subseteq [J]$. We use $|\frakA|$ to denote the cardinality of a set $\frakA$. For a continuously differentiable function $f(x)$, let $\dot f(x)= \rmd f(x)/\rmd x$. In addition, for a bivariate copula function $\bbmC(u_1,u_2)$, let $\dot\bbmC(u_1|u_2)=\partial\bbmC(u_1,u_2)/\partial u_2$ and $\dot\bbmC(u_2|u_1)=\partial\bbmC(u_1,u_2)/\partial u_1$.

\section{The vine copula-based approach} \label{sec:mod}

\subsection{Data} \label{subsec:TMIC}

Let $\bfT_{[J-1]}=(T_1,\ldots,T_{J-1})^\top$ denote the $J-1$ ($J\geq3$) nonterminal event times ($[J]=\{1,\ldots,J\}$ for $J\in\bbN^+$), and let $T_J$ denote the terminal event time. Let $A$ denote the time to independent administrative censoring. The observed terminal event time is given by $X_J= T_J\wedge A$ with censoring status indicator $\Delta_J=\bbone(T_J\leq A)$. For $j\in[J-1]$, the nonterminal event time $T_j$ is right-censored by $X_J$, leading to the observed nonterminal event time $X_j=T_j\wedge X_J$ and the corresponding censoring indicator $\Delta_j=\bbone(T_j\leq X_J)$. As a result, censoring of nonterminal event times by the terminal event is informative whenever $T_J$ is dependent on $T_j$.

Let $\bfZ\in\bbR^{d_Z}$ denote a $d_Z$-dimensional vector of observed covariates. For a sample of $n$ independent subjects, the observed data are given by $\calO = \{\calO_i\equiv(X_{1,i},\Delta_{1,i},\ldots,X_{J,i},\Delta_{J,i},\bfZ_i),i\in[n]\}$. We assume the existence of a maximum follow-up time $\overline t<\infty$ and a constant $\underline p>0$ such that $\bbP(X_j>\overline t|\bfZ)>\underline p$ for all $j\in[J]$, and that every element of $\bfZ$ is bounded with probability one. This is a standard regularity condition required to ensure that the subsequent linear predictor, e.g., $\bfbeta^\top \bfZ$, remains bounded away from zero and infinity over the compact parameter space, which guarantees the strict positivity and uniform boundedness of the marginal hazard functions. Furthermore, the uniform boundedness of $\bfZ$ is necessary to satisfy the bracketing entropy conditions required for the functional convergence of the empirical processes \citep[see, e.g.,][]{Andersen1993, vanderVaart1996}.

\subsection{The vine copula-based modeling} \label{subsec:c-vine}

Modeling such data requires a structured approach to capturing dependence among multiple nonterminal event times and a terminal event, while preserving the interpretability and valid estimation of marginal regression models. Different pairs of event times—including pairs between nonterminal events and between each nonterminal event and the terminal event—may exhibit heterogeneous dependence patterns. To address these features, we adopt a copula-based framework that separates marginal and dependence modeling and allows pairwise associations to vary across event types.

By Sklar's theorem \citep{Sklar1959}, a unique $J$-variate survival copula conditional on covariates, $\bbmC_{[J]}:[0,1]^J\mapsto[0,1]$, exists such that the joint survival function of $\bfT_{[J]}$,
\begin{equation} \label{eq:copula-Jvar}
    S_{[J]}(\bft_{[J]}|\bfZ)=\bbmC_{[J]}\{S_1(t_1|\bfZ),\ldots,S_{J-1}(t_{J-1}|\bfZ),S_J(t_J|\bfZ)\}
\end{equation}
where $\{S_j(t_j|\bfZ),j\in[J]\}$ are respective marginal survival functions (referred to as marginals hereafter) of $\{T_j,j\in[J]\}$. Thus, the joint density of $\bfT_{[J]}$ conditional on covariates, denoted by $f_{[J]}(\bft_{[J]}|\bfZ)$, can be expressed as
\begin{equation}
    f_{[J]}(\bft_{[J]}|\bfZ) = \left\{\prod_{j=1}^J f_j(t_j|\bfZ)\right\} \bbmc_{[J]}(\bfu_{[J],\bfZ}), \label{eq:joint-density}
\end{equation}
where $f_j(t_j|\bfZ) = -\dot S_j(t_j|\bfZ)$ is the marginal density of $T_j$ for $j\in[J]$, $u_{j,\bfZ} = S_j(t_j|\bfZ)$, and $\bbmc_{[J]}(u_{1},\ldots,u_{J}) = \partial^J \bbmC(u_{1},\ldots,u_{J})/\partial u_{1} \cdots \partial u_{J}$ is the joint copula density function. Note that $\bbmc_{[J]}$ can be regarded as the joint density function for $\bfU_{[J],\bfZ}=(U_{1,\bfZ},\ldots, U_{J,\bfZ})'$ with $U_{j,\bfZ}=S_j(T_j|\bfZ)\in[0,1]$.

The vine copula decomposes $\bbmc_{[J]}$ into a product of bivariate copula densities, the structure of which can be illustrated using a graph of $J-1$ ordered trees, $\scrV_{[J]}=(\scrT_1,\ldots,\scrT_{J-1})$, where the $k$-th tree $\scrT_k=(\scrN_k,\scrE_k)$ consists of a set of nodes $\scrN_k$ and a set of edges $\scrE_k$, $k\in[J-1]$. Each edge is associated with a bivariate copula. The vine graph satisfies the following conditions \citep{Bedford2002}: (i) $\scrN_1=[J]$; (ii) $\scrN_k=\scrE_{k-1}$, $k=2,\ldots,J-1$; and (iii) the proximity condition: for any edge $e_k\in\scrE_k$, $k=2,\ldots,J-1$, its nodes, i.e., two edges in $\scrE_{k-1}$, share a common node. We denote edge $e_k$ from tree $\scrT_k$ by $(j_1,j_2|\frakD_{e_k})$, where $\frakD_{e_k}$ is the conditioning set of this edge \citep{Czado2022annualrev}; for $\scrT_1$, $(j_1,j_2|\frakD_{e_1}) =(j_1,j_2)$ since $\frakD_{e_1} \equiv \emptyset$.

For multivariate event times under informative censoring, we specify the first tree of the vine to have a star structure, with the terminal event time serving as the root node. This choice reflects the special role of the terminal event, which induces informative censoring for all nonterminal event times. Accordingly, the edges of the first tree are given by $\scrE_1=\{(j,J): j\in[J-1]\}$, representing all nonterminal--terminal pairs. Each edge $(j,J)$ is associated with a bivariate copula $\bbmC_{j,J}$ that characterizes the joint distribution of $(T_j,T_J)$. This construction enables direct and consistent estimation of the marginal model for each nonterminal event time by explicitly accounting for its dependence on the terminal event \citep{Fine2001}. For subsequent trees, the vine structure may be selected based on the study context, such as a $C$-vine or a $D$-vine. Under this specification, the resulting joint copula density is given in Proposition~\ref{prop:den-v}.

\begin{proposition} \label{prop:den-v}
    Consider a vine graph $\scrV_{[J]}=(\scrT_1,\ldots,\scrT_{J-1})$ for $\bfT_{[J]}$ with $J\geq 3$, where $\scrT_1$ has the root node $J$ and is connected to all other nodes. The joint density of $\bfT_{[J]}$ is given in \eqref{eq:joint-density}, where the joint copula density $\bbmc_{[J]}(\bfu_{[J],\bfZ})$ is given by
    \begin{align} \label{eq:den-v-dec}
        \bbmc_{[J]}(\bfu_{[J],\bfZ})= \bbmc_{[J-1]|J}(u_{1|J,\bfZ}, \ldots, u_{J-1|J,\bfZ})\times \prod_{j=1}^{J-1} \bbmc_{j,J}(u_{j,\bfZ}, u_{J,\bfZ}),
    \end{align}
    where $u_{j|J,\bfZ}=\dot\bbmC_{j,J}(u_{j,\bfZ}|u_{J,\bfZ})$, $j\in[J-1]$, with $\bbmC_{j,J}$ being a bivariate copula for the joint survival function of $(T_j,T_J)$ conditional on covariates, and $\bbmc_{[J-1]|J}$ is a $(J-1)$-variate vine copula density following a user-specified vine structure.
\end{proposition}

Figure~\ref{fig:vg-5} presents two examples of vine graphs for $J=5$ event times. In both examples, the first tree $\calT_1$ has the same star structure with the terminal event time as the root node, while the remaining trees differ in their configurations. Specifically, $(\calT_2,\calT_3,\calT_4)$ form a $C$-vine in Figure~\ref{fig:vg-5}(a) and a $D$-vine in Figure~\ref{fig:vg-5}(b). These examples illustrate how the conditional dependence structure among nonterminal event times can be flexibly specified through different choices of higher-order trees, while preserving the same nonterminal–terminal dependence structure in the first tree. For the $C$-vine structure in Figure~\ref{fig:vg-5}(a), the joint copula density is given by
\begin{align*}
    \bbmc_{[5]}(u_{[5]}) &= \bbmc_{1,2|3,4,5} \times \bbmc_{1,3|4,5} \times \bbmc_{2,3|4,5} \times \bbmc_{1,4|5} \times \bbmc_{2,4|5} \times \bbmc_{3,4|5} \displaybreak[0]\\
    &\quad\times \bbmc_{1,5} \times \bbmc_{2,5} \times \bbmc_{3,5} \times \bbmc_{4,5},
\end{align*}
whereas for the $D$-vine structure in Figure~\ref{fig:vg-5}(b), the joint copula density becomes
\begin{align*}
    \bbmc_{[5]}(u_{[5]}) &= \bbmc_{1,4|2,3,5} \times \bbmc_{1,3|2,5} \times \bbmc_{2,4|3,5} \times \bbmc_{1,2|5} \times \bbmc_{2,3|5} \times \bbmc_{3,4|5} \displaybreak[0]\\
    &\quad\times \bbmc_{1,5} \times \bbmc_{2,5} \times \bbmc_{3,5} \times \bbmc_{4,5}.
\end{align*}

\begin{figure}[htbp]
    \centering
    \begin{subfigure}{\textwidth}
        \resizebox{0.83\columnwidth}{!}{
		\begin{tikzpicture}[
		squarenode1/.style={rectangle, draw=MidnightBlue!60, fill=MidnightBlue!5, very thick, rounded corners=.25cm, minimum height=1cm, minimum width=1cm}
		]

		\node[squarenode1] (T5) at (0,6) {$5$};
		\node[squarenode1] (T4) at (3/1.414,6+3/1.414) {$4$};
		\node[squarenode1] (T3) at (3/1.414,6-3/1.414) {$3$};
        \node[squarenode1] (T2) at (-3/1.414,6-3/1.414) {$2$};
        \node[squarenode1] (T1) at (-3/1.414,6+3/1.414) {$1$};

		\node[squarenode1] (T4T5) at (10,6) {$4,5$};
		\node[squarenode1] (T3T5) at (10+1.5,6+1.5*1.732) {$3,5$};
		\node[squarenode1] (T2T5) at (10+1.5,6-1.5*1.732) {$2,5$};
        \node[squarenode1] (T1T5) at (10-3,6) {$1,5$};

        \node[squarenode1] (T3T4T5) at (0,1) {$3,4|5$};
		\node[squarenode1] (T2T4T5) at (-3,1) {$2,4|5$};
		\node[squarenode1] (T1T4T5) at (3,1) {$1,4|5$};

        \node[squarenode1] (T1T3T4T5) at (7.5,1) {$1,3|4,5$};
		\node[squarenode1] (T2T3T4T5) at (11.5,1) {$2,3|4,5$};

        \node[text width=1.2cm] at (-4.5,6) {\large $\scrT_1$:};
        \node[text width=1.2cm] at (5.5,6) {\large $\scrT_2$:};
        \node[text width=1.2cm] at (-4.5,1) {\large $\scrT_3$:};
        \node[text width=1.2cm] at (5.5,1) {\large $\scrT_4$:};

		\draw[>=latex, -, very thick, MidnightBlue, densely dashed] (T1) -- (T5) node [midway, sloped, above, black] (copula15) {$\bbmC_{1,5}$} node [midway, sloped, below, black] (15) {$1,5$};
		\draw[>=latex, -, very thick, MidnightBlue, densely dashed] (T2) -- (T5) node [midway, sloped, above, black] (copula25) {$\bbmC_{2,5}$} node [midway, sloped, below, black] (25) {$2,5$};
        \draw[>=latex, -, very thick, MidnightBlue, densely dashed] (T3) -- (T5) node [midway, sloped, above, black] (copula35) {$\bbmC_{3,5}$} node [midway, sloped, below, black] (35) {$3,5$};
        \draw[>=latex, -, very thick, MidnightBlue, densely dashed] (T4) -- (T5) node [midway, sloped, above, black] (copula45) {$\bbmC_{4,5}$} node [midway, sloped, below, black] (45) {$4,5$};

		\draw[>=latex, -, very thick, MidnightBlue, densely dashed] (T1T5) -- (T4T5) node [midway, sloped, above, black] (copula14|5) {$\bbmC_{1,4|5}$} node [midway, sloped, below, black] (14|5) {$1,4|5$};
		\draw[>=latex, -, very thick, MidnightBlue, densely dashed] (T2T5) -- (T4T5) node [midway, sloped, above, black] (copula24|5) {$\bbmC_{2,4|5}$} node [midway, sloped, below, black] (24|5) {$2,4|5$};
        \draw[>=latex, -, very thick, MidnightBlue, densely dashed] (T3T5) -- (T4T5) node [midway, sloped, above, black] (copula34|5) {$\bbmC_{3,4|5}$} node [midway, sloped, below, black] (34|5) {$3,4|5$};

		\draw[>=latex, -, very thick, MidnightBlue, densely dashed] (T1T4T5) -- (T3T4T5) node [midway, sloped, above, black] (copula13|45) {$\bbmC_{1,3|4,5}$} node [midway, sloped, below, black] (13|45) {$1,3|4,5$};
		\draw[>=latex, -, very thick, MidnightBlue, densely dashed] (T2T4T5) -- (T3T4T5) node [midway, sloped, above, black] (copula23|45) {$\bbmC_{2,3|4,5}$} node [midway, sloped, below, black] (23|45) {$2,3|4,5$};

		\draw[>=latex, -, very thick, MidnightBlue, densely dashed] (T1T3T4T5) -- (T2T3T4T5) node [midway, sloped, above, black] (copula12|345) {$\bbmC_{1,2|3,4,5}$} node [midway, sloped, below, black] (12|345) {$1,2|3,4,5$};

		\end{tikzpicture}
        }
        \caption{The vine graph of $\scrV_{[5]}=(\scrT_1,\scrT_2,\scrT_3,\scrT_4)$, where $(\scrT_2,\scrT_3,\scrT_4)$ forms a $C$-vine.}
    \end{subfigure}
    \\
    \vspace{1cm}
    \begin{subfigure}{\textwidth}
        \resizebox{\columnwidth}{!}{
		\begin{tikzpicture}[
		squarenode1/.style={rectangle, draw=MidnightBlue!60, fill=MidnightBlue!5, very thick, rounded corners=.25cm, minimum height=1.1cm, minimum width=1.1cm}
		]

		\node[squarenode1] (T5) at (0,6) {$5$};
		\node[squarenode1] (T4) at (3/1.414,6+3/1.414) {$4$};
		\node[squarenode1] (T3) at (3/1.414,6-3/1.414) {$3$};
        \node[squarenode1] (T2) at (-3/1.414,6-3/1.414) {$2$};
        \node[squarenode1] (T1) at (-3/1.414,6+3/1.414) {$1$};

		\node[squarenode1] (T4T5) at (16,6) {$4,5$};
		\node[squarenode1] (T3T5) at (13,6) {$3,5$};
		\node[squarenode1] (T2T5) at (10,6) {$2,5$};
        \node[squarenode1] (T1T5) at (7,6) {$1,5$};

        \node[squarenode1] (T2T3T5) at (0,1) {$2,3|5$};
		\node[squarenode1] (T1T2T5) at (-3,1) {$1,2|5$};
		\node[squarenode1] (T3T4T5) at (3,1) {$3,4|5$};

        \node[squarenode1] (T1T3T2T5) at (7.5,1) {$1,3|2,5$};
		\node[squarenode1] (T2T4T3T5) at (11.5,1) {$2,4|3,5$};

        \node[text width=1.2cm] at (-4.5,6) {\large $\scrT_1$:};
        \node[text width=1.2cm] at (5.5,6) {\large $\scrT_2$:};
        \node[text width=1.2cm] at (-4.5,1) {\large $\scrT_3$:};
        \node[text width=1.2cm] at (5.5,1) {\large $\scrT_4$:};

		\draw[>=latex, -, very thick, MidnightBlue, densely dashed] (T1) -- (T5) node [midway, sloped, above, black] (copula15) {$\bbmC_{1,5}$} node [midway, sloped, below, black] (15) {$1,5$};
		\draw[>=latex, -, very thick, MidnightBlue, densely dashed] (T2) -- (T5) node [midway, sloped, above, black] (copula25) {$\bbmC_{2,5}$} node [midway, sloped, below, black] (25) {$2,5$};
        \draw[>=latex, -, very thick, MidnightBlue, densely dashed] (T3) -- (T5) node [midway, sloped, above, black] (copula35) {$\bbmC_{3,5}$} node [midway, sloped, below, black] (35) {$3,5$};
        \draw[>=latex, -, very thick, MidnightBlue, densely dashed] (T4) -- (T5) node [midway, sloped, above, black] (copula45) {$\bbmC_{4,5}$} node [midway, sloped, below, black] (45) {$4,5$};

		\draw[>=latex, -, very thick, MidnightBlue, densely dashed] (T1T5) -- (T2T5) node [midway, sloped, above, black] (copula12|5) {$\bbmC_{1,2|5}$} node [midway, sloped, below, black] (12|5) {$1,2|5$};
		\draw[>=latex, -, very thick, MidnightBlue, densely dashed] (T2T5) -- (T3T5) node [midway, sloped, above, black] (copula23|5) {$\bbmC_{2,3|5}$} node [midway, sloped, below, black] (23|5) {$2,3|5$};
        \draw[>=latex, -, very thick, MidnightBlue, densely dashed] (T3T5) -- (T4T5) node [midway, sloped, above, black] (copula34|5) {$\bbmC_{3,4|5}$} node [midway, sloped, below, black] (34|5) {$3,4|5$};

		\draw[>=latex, -, very thick, MidnightBlue, densely dashed] (T1T2T5) -- (T2T3T5) node [midway, sloped, above, black] (copula13|25) {$\bbmC_{1,3|2,5}$} node [midway, sloped, below, black] (13|25) {$1,3|2,5$};
		\draw[>=latex, -, very thick, MidnightBlue, densely dashed] (T2T3T5) -- (T3T4T5) node [midway, sloped, above, black] (copula24|35) {$\bbmC_{2,4|3,5}$} node [midway, sloped, below, black] (24|35) {$2,4|3,5$};

		\draw[>=latex, -, very thick, MidnightBlue, densely dashed] (T1T3T2T5) -- (T2T4T3T5) node [midway, sloped, above, black] (copula14|235) {$\bbmC_{1,4|2,3,5}$} node [midway, sloped, below, black] (14|235) {$1,4|2,3,5$};

		\end{tikzpicture}
        }
        \caption{The vine graph of $\scrV_{[5]}=(\scrT_1,\scrT_2,\scrT_3,\scrT_4)$, where $(\scrT_2,\scrT_3,\scrT_4)$ forms a $D$-vine.}
    \end{subfigure}
	\caption{Two examples of vine graphs for $\bfT_{[5]}$, $\scrV_{[5]}=(\scrT_1,\scrT_2,\scrT_3,\scrT_4)$. The first tree $\scrT_1$ is the same for the two graphs, but the remaining three trees, $(\scrT_2,\scrT_3,\scrT_4)$, form a $C$-vine in panel (a), and a $D$-vine in panel (b). Associated bivariate copulas are given above the corresponding edges.}
	\label{fig:vg-5}
\end{figure}

To evaluate each bivariate copula density in the vine decomposition, its arguments are expressed as conditional survival functions that are computed recursively from lower-order trees.  For an edge $e_k=(j_1,j_2 | \frakD_{e_k})$ in tree $\scrT_k$, the associated bivariate copula is evaluated at the conditional survival functions
\begin{align*}
    u_{j | \frakD_{e_k},\bfZ} = \bbP(T_j \ge t_j | \bfT_{\frakD_{e_k}}=\bft_{\frakD_{e_k}}, \bfZ), \quad j=j_1,j_2.
\end{align*}
These conditional survival functions are obtained from the parent edges of $e_k$ in $\scrT_{k-1}$. Suppose $e_k$ has parent edges $(j_1,j^* | \frakD_{e_k}\setminus\{j^*\})$ and $(j_2,j^* | \frakD_{e_k}\setminus\{j^*\})$, where $j^* \in \frakD_{e_k}$. Then, for $j=j_1,j_2$,
\begin{align*}
    u_{j | \frakD_{e_k},\bfZ} = \dot\bbmC_{j,j^* | \frakD_{e_k}\setminus\{j^*\}} \left(u_{j | \frakD_{e_k}\setminus\{j^*\},\bfZ} \middle| u_{j^* | \frakD_{e_k}\setminus\{j^*\},\bfZ}\right).
\end{align*}
Take $\bbmc_{1,3 | 4,5}$ as an example. It is associated with edge $e_3=(1,3|4,5)$ in tree $\scrT_3$ of the $C$-vine in Figure~\ref{fig:vg-5}(a). Its two arguments are $u_{1|4,5}=\dot\bbmC_{1,4|5}(u_{1|5}| u_{4|5})$ and $u_{3|4,5}=\dot\bbmC_{3,4|5}(u_{3|5}| u_{4|5})$, where $(1,4|5)$ and $(3,4|5)$ are the parent edges of $e_3$ in tree $\scrT_2$.

{
For higher-order trees, recursively evaluating these conditional survival functions requires tracking the ancestor edges of each edge in the vine. To formalize this recursive structure, we define a parent-mapping function $\xi_m: \calE_{m+1} \to 2^{\calE_m}$ for $m \in \{1, \dots, J-2\}$, which maps an edge $e \in \calE_{m+1}$ to its set of parent edges in $\calT_m$. For any two set-valued maps $F: X \to 2^Y$ and $G: Y \to 2^Z$, their composition $G \circ F: X \to 2^Z$ is defined by $(G \circ F)(x) = \cup_{y \in F(x)} G(y)$. Let $\overline{\calE}_r(e_k) \subset \calE_{k-r}$ denote the subset of ancestor edges of $e_k$ in tree $\calT_{k-r}$, for $1 \leq r \leq k-1$. This subset is generated by the recursive composition of the parent-mapping functions:
\begin{align*}
    \overline{\calE}_r(e_k) = (\xi_{k-r} \circ \xi_{k-r+1} \circ \dots \circ \xi_{k-1})(e_k).
\end{align*}
Continuing the previous example, for edge $e_3=(1,3|4,5)$, $\overline{\calE}_1(e_3) = \{(1,4|5), (3,4|5)\}$ (ancestor edges in $\scrT_2$) and $\overline{\calE}_2(e_3) = \{(1,5), (3,5), (4,5)\}$ (ancestor edges in $\scrT_1$).

Finally, for an edge $e_k = (j_1, j_2|\frakD_{e_k})$, let
$\frakn(e_k) = \{j_1, j_2\} \cup \frakD_{e_k} \subset [J]$ denote the indices
of the event times involved in $e_k$. For example, $\frakn(e_3) = \{1,3,4,5\}$
for $e_3 = (1,3|4,5)$. The following Corollary~\ref{coro:den-vk} gives the joint density of $T_{\frakn(e_k)}$.

\begin{corollary} \label{coro:den-vk}
    The joint density of $\bfT_{\frakn(e_k)}$ conditional on $\bfZ$ is
    \begin{align*}
        f_{\frakn(e_k)}(t_{\frakn(e_k)} |\bfZ) = \bbmc_{\frakn(e_k)}(u_{\frakn(e_k), \bfZ}) \prod_{j \in \frakn(e_k)} f_j(t_j | \bfZ)
    \end{align*}
    where the joint copula density is:
    \begin{align} \label{eq:den-vk}
        \bbmc_{\frakn(e_k)}(u_{\frakn(e_k), \bfZ}) = \bbmc_{e_k} \prod_{r=1}^{k-1} \prod_{e \in \overline{\calE}_r(e_k)} \bbmc_e.
    \end{align}
\end{corollary}

Applying Corollary~\ref{coro:den-vk} to $e_3 = (1,3|4,5)$ yields the joint density function of $(T_1,T_3,T_4,T_5)$, given by $f_{1,3,4,5}(t_1,t_3,t_4,t_5|\bfZ) = \bbmc_{1,3,4,5}(u_{1,\bfZ},u_{3,\bfZ},u_{4,\bfZ},u_{5,\bfZ})\times f_1(t_1|\bfZ)\times\allowbreak f_3(t_3|\bfZ) f_4(t_4|\bfZ)f_5(t_5|\bfZ)$, where $\bbmc_{1,3,4,5} = \bbmc_{1,3|4,5} \times \bbmc_{1,4|5} \times \bbmc_{3,4|5} \times \bbmc_{1,5} \times \bbmc_{3,5}\times \bbmc_{4,5}$.
}

\begin{remark} \label{rmk:identify}
    The identifiability issue in semi-competing risks data has been discussed in previous works \citep{Fine2001, peng2007regression, Chen2012, Li2023}. The bivariate copula $\bbmC_{j,J}$ specifies the joint density of $(T_j,T_J)$ on the region where $T_j\leq T_J$ for $j\in[J-1]$, only on which the joint density is nonparametrically identifiable. Thus, we label $\frakR_{[J]} = \{\bft_{[J]}: t_1,\ldots, t_{J-1} \in(0,t_J]\}$ the identifiable region of $\bfT_{[J]}$, and the joint density of $\bfT_{[J]}$ on $\frakR_{[J]}$ is specified by the joint density in \eqref{eq:den-v-dec}.
\end{remark}

\begin{remark}
    The copula function $\bbmC_{[J]}$ was originally introduced to connect the joint cumulative distribution function (CDF) of multivariate random variables with the marginal CDFs. The survival copula considered in this article shares properties similar to those of the original copula \citep{Nelsen2006}, connecting the joint survival function with the marginals.
\end{remark}

\begin{remark}\label{rmk:simplifying}
{
The proposed vine copula framework adopts the simplifying assumption of vine copulas \citep[e.g.,][]{stober2013, Nagler2025}. Under this assumption, the realized values of the conditioning event times enter each conditional bivariate copula $\bbmC_{j_1,j_2|\frakD}$ only through the conditional survival functions $u_{j_1|\frakD,\bfZ}$ and $u_{j_2|\frakD,\bfZ}$; the copula family and parameter $\bfgamma_{j_1,j_2|\frakD}$ do not depend on these realized values.
}
\end{remark}

\subsection{Model specification} \label{subsec:mod-spec}

{
To assess the direct covariate effects on the marginals, we specify $S_j(t_j|\bfZ)$ using a Cox proportional hazards (PH) model \citep{cox1972regression},
\begin{equation} \label{eq:marginal}
    S_j(t_j|\bfZ;\bftheta_j) = \exp\left\{-\int_0^{t_j}e^{\bfbeta_j^\top\bfZ}\rmd \Lambda_j(s_j)\right\}.
\end{equation}
The marginal model parameters are $\bftheta_j\equiv\{\bfbeta_j,\Lambda_j(\cdot)\}$, where $\bfbeta_j$ is a $d_Z$-dimensional vector of regression coefficients, and $\Lambda_j(\cdot)$ is an unspecified baseline cumulative hazard function, assumed to be increasing and continuously differentiable.

Each bivariate copula in \eqref{eq:den-v-dec} is specified using a parametric copula family, e.g., Archimedean or elliptical; expressions of some popular families are provided in Supp. Section S1. The copula parameter associated with $\bbmc_{j_1,j_2|\frakD_{e_k}}$, denoted by $\alpha_{j_1,j_2|\frakD_{e_k},\bfZ}$, might depend on $\bfZ$. It can be modeled by
\begin{equation}
\alpha_{j_1,j_2|\frakD_{e_k},\bfZ}=g_{j_1,j_2|\frakD_{e_k}}(\bfgamma_{j_1,j_2|\frakD_{e_k}}^\top\widetilde\bfZ), \label{eq:alpha}
\end{equation}
where $\widetilde\bfZ = (1,\bfZ^\top)^\top$, $\bfgamma_{j_1,j_2|\frakD_{e_k}}\in\bbR^{d_Z+1}$ is a $(d_Z+1)$-dimensional event-pair-specific parameter vector, and $g_{j_1,j_2|\frakD_{e_k}}(\cdot)$ is a pre-specified link function in conjunction with the selected copula family \citep{Vatter2018}. Examples of $g_{j_1,j_2|\frakD_{e_k}}(\cdot)$ include $g_{j_1,j_2|\frakD_{e_k}}(x)=e^x$ for Clayton, $g_{j_1,j_2|\frakD_{e_k}}(x)=x$ for Frank, $g_{j_1,j_2|\frakD_{e_k}}(x)=e^x+1$ for Gumbel, and $g_{j_1,j_2|\frakD_{e_k}}(x)=(e^{2x}-1)/(e^{2x}+1)$ for Gaussian. Because of the one-to-one correspondence between $\alpha_{j_1,j_2|\frakD_{e_k},\bfZ}$ and $\bfgamma_{j_1,j_2|\frakD_{e_k}}$, we hereafter write $\bbmC_{j_1,j_2|\frakD_{e_k}}(\cdot;\alpha_{j_1,j_2|\frakD_{e_k},\bfZ})$ as $\bbmC_{j_1,j_2|\frakD_{e_k}}(\cdot;\bfgamma_{j_1,j_2|\frakD_{e_k}})$.

{
\begin{remark}\label{rmk:covariate}
If the application requires modeling complex covariate effect heterogeneity, the design vector $\bfZ$ can be directly augmented with polynomial basis functions or interaction terms without altering the underlying copula structure. In addition, although we use a common covariate vector $\bfZ$ in both the marginal models and the copula parameter models for notational simplicity, the covariates entering each model component need not be identical; for example, different subsets of $\bfZ$ may be used for different marginals or copula parameters.
\end{remark}
}

\subsection{The full likelihood}

We present the full likelihood function under the vine copula-based approach, which depends on the censoring status indicator for each event time. For a non-empty subset $\frakJ \subseteq [J]$, define its observed index set $\frako_{\frakJ,i} =\{j|j\in \frakJ, \Delta_{j,i}=1\}$ and censored index set $\frakc_{\frakJ,i}=\{j|j\in \frakJ, \Delta_{j,i}=0\}$. Let $U_{j,\bfZ,i}=S_j(X_{j,i}|\bfZ_i;\bftheta_j)$, $j \in [J]$, often referred to as the pseudo observations, and define $\bfU_{\frakJ,\bfZ,i}^{\frako}=\{U_{j,\bfZ,i},j\in \frako_{\frakJ,i}\}$ and $\calU_{\frakJ,\bfZ,i}^\frakc=\prod_{j\in \frakc_{\frakJ,i}}[0,U_{j,\bfZ,i}]$. Observed event times enter the full likelihood function via $\bfU_{[J],\bfZ,i}^{\frako}$, while censored event times are integrated over their corresponding range $\calU_{[J],\bfZ,i}^\frakc$. For ease of presentation, we rearrange the order of the arguments of the joint copula density $\bbmc_{[J]}$ by their censoring statuses, $\bbmc_{[J]}(\bfU_{[J],\bfZ,i}^\frako,\bfu_{[J],\bfZ,i}^{\frakc})$, where $\bfu_{[J],\bfZ,i}^{\frakc}$ is a set of $|\frakc_{[J],i}|$ variables associated with the integral. Let $\bfOmega$ denote the collection of all model parameters, and the log-likelihood function for subject $i$ is
\begin{align} \label{eq:log-lik-all}
    \ell(\bfOmega;\calO_i) = \sum_{j \in [J]} \Delta_{j,i}\log\left\{-f_j(X_{j,i}|\bfZ_i;\bftheta_j)\right\} + h(\bfOmega;\bfU_{[J],\bfZ,i},\bfDelta_{[J],i}),
\end{align}
where
\begin{align} \label{eq:log-lik-copula}
    h(\bfOmega;\bfU_{[J],\bfZ,i},\bfDelta_{[J],i}) = \log\int_{\calU_{[J],\bfZ,i}^\frakc}\bbmc_{[J]}\left(\bfU_{[J],\bfZ,i}^\frako,\bfu_{[J],\bfZ,i}^{\frakc}\right)\rmd \bfu_{[J],\bfZ,i}^{\frakc}.
\end{align}

We also provide an alternative construction of the full likelihood function using a counting process representation in Supp. Section S2.

\subsection{Comparison with multivariate and nested copulas}\label{sec:model-comp}

{

\citet{Li2020} modeled the joint survival function of all event times using a single parametric multivariate Archimedean copula, thereby inducing homogeneous pairwise dependence among them. \citet{Li2023} extended this framework by adopting a two-level nested copula, in which the joint survival function of the nonterminal event times, $S_{[J-1]}(\bft_{[J-1]}| \bfZ)$, is first modeled by a $(J-1)$-variate copula and then linked to the terminal event survival function $S_J(t_J | \bfZ)$ through a second-level bivariate Archimedean copula. This hierarchical construction allows the dependence between the nonterminal and terminal event times to differ from that among the nonterminal event times, while maintaining homogeneous pairwise dependence within the nonterminal event times and a common association between each nonterminal event time and the terminal event.

These two works directly characterize the unconditional dependence among nonterminal event times while accounting for informative censoring by the terminal event time. By contrast, our proposed vine copula framework represents the joint density function through a sequence of bivariate copulas arranged according to a vine structure, with the higher-order trees directly characterizing conditional dependence among the nonterminal event times. This pair-copula construction allows distinct event pairs to be modeled using different copula families (e.g., Archimedean or elliptical) and corresponding dependence parameters. The flexibility comes with the simplifying assumption for vine copulas (Remark~\ref{rmk:simplifying}), under which the copula family and its parameter remain invariant to the realized values of the conditioning event times. In contrast, the nested copula framework does not rely on the simplifying assumption. Instead, it is governed by the sufficient nesting condition \citep{Hofert2013}, which imposes a different set of structural constraints to ensure a valid joint distribution.

The proposed vine copula framework and the multivariate and nested copula frameworks of \citet{Li2020} and \citet{Li2023} should therefore be viewed as distinct, complementary model classes for multivariate event-time data under informative censoring. They differ in the construction of the joint survival function, the structural assumptions imposed on the dependence model, and the dependence summaries (e.g., Kendall's $\tau$) that naturally arise from their respective formulations. These dependence summaries have distinct scientific interpretations, which are discussed further in Section~\ref{sec:disc}.
}

\section{Stage-wise estimation and inference} \label{sec:est}

\subsection{Stage-wise estimation} \label{sec:estimation}

We estimate $\bfOmega$ via a stage-wise procedure. Let $\bfOmega_{\scrT_k}$ denote model parameters associated with tree $\scrT_k$ for $k\in[J-1]$, where $\bfOmega_{\scrT_1}=\{\bftheta_j,j\in[J]\}\cup\{\bfgamma_{j,J},j\in[J-1]\}$, and $\bfOmega_{\scrT_k}=\{\bfgamma_{e_k},e_k\in\scrE_k\}$ for $k=2,\ldots,J-1$. We define $\calO_j=\{\calO_{j,i}\equiv(X_{j,i},\Delta_{j,i},\bfZ_i),i\in[n]\}$ as the observed data related to event $j\in[J]$, and the stage-wise estimation procedure is given as follows.

\paragraph{Stage 1: Estimate $\bfOmega_{\scrT_1}$.} The estimation of $\bfOmega_{\scrT_1}$ consists of two steps. At the first step, we obtain $\widehat \bftheta_J \equiv\{\widehat\bfbeta_J,\widehat\Lambda_J(\cdot)\}$, the estimate of model parameters in $S_J(t_J|\bfZ,\bftheta_J)$, with observed data $\calO_J$ alone, since $T_J$ is only subject to independent censoring. Specifically, the baseline function $\Lambda_J(\cdot)$ is regarded as a non-decreasing step function that jumps only at the observed terminal event times. Let $0<t_{J,1} < t_{J,2} < \cdots < t_{J,\kappa_J}\leq\overline t$ denote the ordered observed terminal event times, and let $\rmd \Lambda_{J,l}>0$ denote the jump size of $\Lambda_J(\cdot)$ at time $t_{J,l}$, for $l\in[\kappa_J]$. The terminal marginal parameters $\bftheta_J$ can be re-expressed as $(\bfbeta_J^\top, \rmd\Lambda_{J,1},\ldots, \rmd\Lambda_{J,\kappa_J})^\top$. The estimate $\widehat\bftheta_J$ is obtained via maximizing the log-likelihood function $\sum_{i=1}^n \ell_J(\bftheta_J;\calO_{J,i})$. {Under the Cox PH model in \eqref{eq:marginal},
\begin{align*}    &\ell_J(\bftheta_J;\calO_{J,i}) = -\int_0^{\overline t} e^{\bfbeta_J^\top\bfZ_i}\rmd\Lambda_J(s) + \Delta_{J,i}\left\{ \log\rmd\Lambda_J(X_{J,i}) + \bfbeta_J^\top\bfZ_i \right\}.
\end{align*}
The resulting estimator $\widehat\bftheta_J$ corresponds to the Cox maximum partial likelihood estimator for $\bfbeta_J$ together with the Breslow estimator for $\Lambda_J$.
}

At the second step, we {\it separately} estimate $\bfOmega_{j,J}\equiv\{\bftheta_j,\bfgamma_{j,J}\}$ for each $j\in[J-1]$ via the PMLE method in \citet{Arachchige2024} proposed for semi-competing risks data. Similar to the first step, let $0<t_{j,1} < t_{j,2} < \cdots < t_{j,\kappa_j}\leq\overline t$ denote the ordered observed event times for the nonterminal event $j\in[J-1]$, and let $\rmd \Lambda_{j,l}>0$ denote the jump size of $\Lambda_j(\cdot)$ at time $t_{j,l}$, for $l\in[\kappa_j]$. The nonterminal marginal parameters $\bftheta_j$ can be re-expressed as $(\bfbeta_j^\top, \rmd\Lambda_{j,1},\ldots, \rmd\Lambda_{j,\kappa_j})^\top$. For each $j\in[J-1]$, with observed data of subject $i$, $\calO_{(j,J),i} =\calO_{j,i}\cup \calO_{J,i}$, the complete log-likelihood function is $\ell_{j,J}^{\mathrm{comp}}(\bfOmega_{j,J},\bftheta_J;\calO_{(j,J),i}) = \ell_{j,J}(\bfOmega_{j,J};U_{J,\bfZ,i},\calO_{(j,J),i}) + \widetilde{\ell}_J(\bftheta_J;\calO_{J,i})$. {Under the Cox PH model in \eqref{eq:marginal}, $\widetilde{\ell}_J(\bftheta_J;\calO_{J,i}) = \Delta_{J,i}\left\{-\Lambda_{J,i}+\log \rmd\Lambda_J(X_{J,i}) + \bfbeta_J^\top\bfZ_i\right\}$},
a function of $\bftheta_J$ alone, and
{\begin{align} \label{eq:log-lik-j-J}
    & \ell_{j,J}(\bfOmega_{j,J};U_{J,\bfZ,i},\calO_{(j,J),i}) = \Delta_{j,i}\Delta_{J,i} \log \bbmc_{j,J}(U_{j,\bfZ,i},U_{J,\bfZ,i};\bfgamma_{j,J}) \displaybreak[0]\nonumber\\
    &\quad+ \Delta_{j,i}(1 - \Delta_{J,i}) \log \dot \bbmC_{j,J}(U_{J,\bfZ,i}|U_{j,\bfZ,i};\bfgamma_{j,J}) \displaybreak[0]\nonumber\\
    &\quad+ (1-\Delta_{j,i})\Delta_{J,i} \log \dot \bbmC_{j,J}(U_{j,\bfZ,i}|U_{J,\bfZ,i};\bfgamma_{j,J}) \displaybreak[0]\nonumber\\
    &\quad+ (1-\Delta_{j,i})(1-\Delta_{J,i}) \log \bbmC_{j,J}(U_{j,\bfZ,i},U_{J,\bfZ,i};\bfgamma_{j,J}) \displaybreak[0]\nonumber\\
    &\quad+ \Delta_{j,i} \left\{-\int_0^{X_{j,i}} e^{\bfbeta_j^\top\bfZ_i}\rmd\Lambda_j(s)+\log\rmd\Lambda_j(X_{j,i}) + \bfbeta_j^\top\bfZ_i \right\},
\end{align} which depends on $\bftheta_J$ only through $U_{J,\bfZ,i}$.} The estimate $\widehat\bfOmega_{j,J}=\{\bfthetahat_j,\bfgammahat_{j,J}\}$ is obtained via maximizing the pseudo log-likelihood function $\sum_{i=1}^n \ell_{j,J}(\bfOmega_{j,J};\Uhat_{J,\bfZ,i},\allowbreak\calO_{(j,J),i})$, where $\Uhat_{J,\bfZ,i}=S_J(X_{J,i}|\bfZ_i,\bfthetahat_J)$. Since $\widetilde\ell_J$ does not involve $\bfOmega_{j,J}$, the estimate $\widehat\bfOmega_{j,J}=\{\bfthetahat_j,\bfgammahat_{j,J}\}$, equivalently, maximizes $\sum_{i=1}^n \allowbreak\ell_{j,J}^{\mathrm{comp}}(\bfOmega_{j,J},\bfthetahat_J,\calO_{(j,J),i})$, where $\bftheta_J$ is regarded as the nuisance parameter.

\paragraph{Stage $k$: Estimate $\bfOmega_{\scrT_k}$, $k=2,\ldots,J-1$.} At stage $k$, we estimate each bivariate copula parameter $\bfgamma_{e_k}$ separately for edge $e_k \in \scrE_k$. Recall that $\frakn(e_k)$ collects the indices of the event times involved in this edge, and $\overline{\scrE}(e_k)\equiv\{\overline{\scrE}_{r}(e_k),r\in[k-1]\}$ consists of the edges of all previous trees that are the ancestor nodes of $e_k$. Let $\bfthetahat_{\frakn(e_k)}=\{\bfthetahat_j,j\in \frakn(e_k)\}$ and $\bfUhat_{\frakn(e_k),\bfZ,i}=\{\Uhat_{j,\bfZ,i}, j\in \frakn(e_k)\}$, estimated pseudo-observations obtained from Stage 1, and $\bfGammahat_{e_k}=\{\bfgammahat_{e}: e \in \overline{\scrE}(e_k)\}$, estimated copula parameters from previous trees. With observed data $\calO_{\frakn(e_k)} = \cup_{j \in \frakn(e_k)}\calO_j=\{\calO_{\frakn(e_k),i},i\in[n]\}$, the pseudo-log-likelihood function for estimating $\bfgamma_{e_k}$ is given by $\sum_{i=1}^n \ell_{e_k}(\bfgamma_{e_k};\bfUhat_{\frakn(e_k),\bfZ,i},\bfGammahat_{e_k},\calO_{\frakn(e_k),i})$,
with
\begin{align*}
    &\ell_{e_k}(\bfgamma_{e_k};\bfUhat_{\frakn(e_k),\bfZ,i},\bfGammahat_{e_k},\calO_{\frakn(e_k),i}) \displaybreak[0]\\
    &= \log \int_{\calUhat_{\frakn(e_k),\bfZ,i}^\frakc}\bbmc_{\frakn(e_k)}\left(\bfUhat_{\frakn(e_k),\bfZ,i}^\frako,\bfu_{\frakn(e_k),\bfZ,i}^{\frakc}\right)\rmd \bfu_{\frakn(e_k),\bfZ,i}^{\frakc},
\end{align*}
where $\bbmc_{\frakn(e_k)}$ is given in \eqref{eq:den-vk}. The nuisance parameters in $\ell_{e_k}$ are $\bftheta_{\frakn(e_k)}=\{\bftheta_j: j\in \frakn(e_k)\}$ and $\bfGamma_{e_k}=\{\bfgamma_e: e \in \overline{\scrE}(e_k)\}$.

In Table~\ref{tab:ill-est}, we demonstrate the above stage-wise PMLE procedure by listing the to-be-estimated parameters at each stage and the corresponding estimates from previous stages that are used in the current estimation. We consider the case of $J=4$, i.e., with three nonterminal events. Note that the $C$-vine structure of $\bbmc_{1,2,3|4}$ is the same as the $D$-vine in this scenario, and an illustration of the vine graph is shown in Web Figure S1 in Supp. Section S3. Compared to the MLE obtained by maximizing the full likelihood given in \eqref{eq:log-lik-all} with respect to all parameters simultaneously, our stage-wise PMLE is less efficient but more convenient to implement in practice, with lower computational cost.

\begin{table}[htbp]
    \centering
    \small
    \begin{tabular}{l l l}
    \toprule
    Stages & Parameters to be estimated & Previous estimates used \\
    \midrule
    Stage 1 & $\{\bfbeta_4,\Lambda_4(\cdot)\}$ & $-$\\
    & $\{\bfbeta_1,\Lambda_1(\cdot),\bfgamma_{1,4}\}$ & $\bfUhat_{4,\bfZ}$\\
    & $\{\bfbeta_2,\Lambda_2(\cdot),\bfgamma_{2,4}\}$ & $\bfUhat_{4,\bfZ}$ \\
    & $\{\bfbeta_3,\Lambda_3(\cdot),\bfgamma_{3,4}\}$ & $\bfUhat_{4,\bfZ}$\\
    \midrule
    Stage 2 & $\bfgamma_{1,3|4}$ & $\{\bfUhat_{1,\bfZ},\bfUhat_{3,\bfZ}, \bfUhat_{4,\bfZ}, \bfgammahat_{1,4}, \bfgammahat_{3,4}\}$ \\
    & $\bfgamma_{2,3|4}$ & $\{\bfUhat_{2,\bfZ},\bfUhat_{3,\bfZ}, \bfUhat_{4,\bfZ}, \bfgammahat_{2,4}, \bfgammahat_{3,4}\}$\\
    \midrule
    Stage 3 & $\bfgamma_{1,2|3,4}$ & $\{\bfUhat_{1,\bfZ}, \bfUhat_{2,\bfZ},\bfUhat_{3,\bfZ}, \bfUhat_{4,\bfZ}, \bfgammahat_{1,4}, \bfgammahat_{2,4}, \bfgammahat_{3,4}, \bfgammahat_{1,3|4}, \bfgammahat_{2,3|4}\}$ \\
    \bottomrule
    \end{tabular}
    \caption{An illustrative example of the stage-wise estimation procedure with $J=4$.}\label{tab:ill-est}
\end{table}

\begin{remark} \label{rmk:pmle-modify}
    The stage-wise PMLE can be modified in conjunction with variations in model specification. For example, when some model component specifications are identical, we can estimate the shared parameters from a pooled pseudo-log-likelihood function. Specifically, in Simulation Study III (in Supp. Section S5.3) with $J=3$ and no covariates, we assume $\bbmC_{1,3}$ and $\bbmC_{2,3}$ are both Clayton with the same parameter $\alpha_{1,3}=\alpha_{2,3}=\alpha$, i.e., homogeneous nonterminal-terminal dependence. Under this assumption, at the first stage, the shared copula parameter $\alpha$ along with marginal parameters $\{\bfbeta_1,\Lambda_1(\cdot),\bfbeta_2,\Lambda_2(\cdot)\}$ are estimated by maximizing a pooled log-likelihood function $\sum_i \{\ell_{1,3}(\alpha,\bfbeta_1,\Lambda_1;\Uhat_{3,\bfZ,i},\calO_{(1,3),i}) + \ell_{2,3}(\alpha,\bfbeta_2,\Lambda_2;\Uhat_{3,\bfZ,i},\calO_{(2,3),i})\}$.
\end{remark}

{
\begin{remark}
The proposed stage-wise PMLE differs from standard sequential estimation procedures for vine copulas under independent censoring, such as the inference functions for margins (IFM) approach of \citet{Barthel2018}. Under independent censoring, the marginal distributions are estimated independently of the copula model. In contrast, informative censoring prevents independent marginal estimation. Consequently, Stage 1 jointly estimates the nonterminal marginal parameters $(\bftheta_j)$ and the corresponding first-tree copula parameters $(\bfgamma_{j,J})$ through pairwise pseudo-likelihoods, thereby accounting for informative censoring before estimating the higher-tree copulas.
\end{remark}
}

\subsection{Theoretical properties}

We present the theoretical properties of the stage-wise estimators ${\widehat\bftheta_j,j\in[J]}$ and ${\widehat\bfgamma_{e_k},e_k\in\scrE_k,k\in[J-1]}$.
The regularity conditions and proofs are provided in Supp. Sections S4.2 and S4.3. Let
$\{\bftheta_j^0\equiv\{\bfbeta_j^0,\Lambda_j^0(\cdot)\},j\in[J]\}$ and
$\{\bfgamma_{e_k}^0,e_k\in\scrE_k,k\in[J-1]\}$
denote the true values of
$\{\bftheta_j,j\in[J]\}$ and
$\{\bfgamma_{e_k},e_k\in\scrE_k,k\in[J-1]\}$, respectively.
We use $\|\cdot\|_{l^\infty[0,\overline t]}$ to denote the supremum norm in $[0,\overline t]$. Let $BV[0,\overline t]$ be the space of real-valued functions with bounded variations in $[0,\overline t]$, and $\|w(\cdot)\|_{BV[0,\overline t]}$ the total variation of a real-valued function $w(t)$ on $[0,\overline t]$. Also, define $\scrL=\{w(t):\|w(\cdot)\|_{BV[0,\overline t]}\leq 1\}$. Then, $\widehat\Lambda_j(\cdot)$, for $j\in[J]$, can be viewed as bounded linear functionals in $l^\infty(\scrL)$. The consistency of the stage-wise estimators is given in Proposition~\ref{prop:PMLE-consistency}.

\begin{proposition} \label{prop:PMLE-consistency}
    {Under conditions (C1) - (C8)} listed in Supp. Section S4.2,
    \begin{itemize}
        \item[(a)] with probability one, $\|\bfbetahat_J-\bfbeta_J^0\|\rightarrow0$ and $\|\Lambdahat_J(\cdot)-\Lambda_J^0(\cdot)\|_{l^\infty[0,\overline t]}\rightarrow0$;
        \item[(b)] for $j\in[J-1]$, with probability one, $\|\bfgammahat_{j,J}-\bfgamma_{j,J}^0\|\rightarrow0$, $\|\bfbetahat_j-\bfbeta_j^0\|\rightarrow0$, and $\|\Lambdahat_j(\cdot)-\Lambda_j^0(\cdot)\|_{l^\infty[0,\overline t]}\rightarrow0$;
        \item[(c)] for $e_k\in\scrE_k$ and $k=2,\ldots,J-1$, with probability one, $\|\bfgammahat_{e_k}-\bfgamma_{e_k}^0\|\rightarrow0$.
    \end{itemize}
\end{proposition}
The consistency of the terminal marginal parameter estimators $\widehat\bftheta_J$ in Proposition~\ref{prop:PMLE-consistency}(a) follows from standard asymptotic results for the Cox PH model; see, for example, \citet{Tsiatis1981} and \citet{AndersenGill1982}. It also follows as a special case of the more general semiparametric transformation-model results of \citet{Zeng2006}. The proof of Proposition~\ref{prop:PMLE-consistency}(b) largely follows the arguments of \citet{Chen2012}, which investigated estimators of $(\bftheta_J, \bfOmega_{j,J})$ that simultaneously maximize the complete log-likelihood $\sum_{i=1}^n \ell^{\mathrm{comp}}_{j,J}(\bfOmega_{j,J},\bftheta_J;\calO_{(j,J),i})$. However, due to the nature of PMLE, the stage-wise estimators require different regularity conditions. The consistency of copula parameter estimators $\bfgammahat_{e_k}$ for $e_k\in\scrE_k$ and $k=2,\ldots,J-1$ in Proposition~\ref{prop:PMLE-consistency}(c) is established following techniques and procedures in \citet{Chen2010joe} for the PMLE of the second-stage copula parameter in  multivariate survival functions. The consistency of the estimators in the later stages depends on that of those in the earlier stages.

We present the asymptotic normality of the stage-wise estimators in Proposition~\ref{prop:PMLE-AN}.

\begin{proposition} \label{prop:PMLE-AN}
    {Under conditions (C1) - (C8) and (A1) - (A2)} listed in Supp. Sections S4.2 and S4.3,
    \begin{itemize}
        \item[(a)] $n^{1/2}\{\widehat\bfbeta_J-\bfbeta_J^0,\widehat\Lambda_J(\cdot)-\Lambda_J^0(\cdot)\}$ converges weakly to a zero-mean Gaussian process in the metric space $\bbR^{d_Z}\times l^\infty(\scrL)$;
        \item[(b)] for $j\in[J-1]$, $n^{1/2}\{\widehat\bfbeta_j-\bfbeta_j^0,\widehat\Lambda_j(\cdot)-\Lambda_j^0(\cdot),\widehat\bfgamma_{j,J}-\bfgamma_{j,J}^0\}$ converges weakly to a zero-mean Gaussian process in $\bbR^{d_Z}\times l^\infty(\scrL)\times\bbR^{d_Z+1}$;
        \item[(c)] for $e_k\in\scrE_k$ and $k=2,\ldots,J-1$, $n^{1/2}(\widehat\bfgamma_{e_k}-\bfgamma_{e_k}^0)$ converges weakly to a zero-mean $(d_Z+1)$-variate normal distribution.
    \end{itemize}
\end{proposition}

Consider the following linear functionals $n^{1/2}[\int_0^{\overline t}\bfa_J^\top(\widehat\bfbeta_J-\bfbeta_J^0)+b_J(t)\{\rmd\widehat\Lambda_J(t)-\rmd\Lambda_J^0(t)\}]$ and $n^{1/2}[\int_0^{\overline t}\bfa_j^\top(\widehat\bfbeta_j-\bfbeta_j^0)+b_j(t)\{\rmd\widehat\Lambda_j(t)-\rmd\Lambda_j^0(t)\}+\bfc_j^\top(\widehat\bfgamma_{j,J}-\bfgamma_{j,J}^0)]$, where $\bfa_J$, $\bfa_j$, and $\bfc_j$ are real vectors, and $b_J(t),b_j(t)\in BV[0,\overline t]$. Also, let $\bfb_J$ and $\bfb_j$ respectively be vectors consisting of the values of $b_J(t)$ and $b_j(t)$ evaluated at the observed times for events $J$ and $j\in[J-1]$. Proposition~\ref{prop:PMLE-AN} implies that these linear functionals are also asymptotically normal with mean zero and variance-covariance matrices $\bfq_J^\top(\bfSigma_{J}^0)^{-1}\bfq_J$ and $\bfq_{j,J}^\top\bfSigma_{(j,J),\bfOmega}^0\bfq_{j,J}$ with $\bfq_J=(\bfa_J^\top,\bfb_J^\top)^\top$ and $\bfq_{j,J}=(\bfa_j^\top,\bfb_j^\top,\bfc_j^\top)^\top$. Here, the variance-covariance matrix $\bfSigma_J^0$ for $\bftheta_J$ is equivalent to the inverse of the Fisher information, defined as $\bfI_{J}^0=-\bbE^0\{\ddot \ell_J(\bftheta_J^0;\calO_{J,i})\}$ with $\ddot \ell_J = \nabla^2_{\bftheta_J}\ell_J$. However, the variance-covariance matrices $\bfSigma_{(j,J),\bfOmega}^0$ for $\bfOmega_{j,J}$ and $\bfSigma_{e_k,\bfgamma}^0$ for $\bfgamma_{e_k}$ are
not equal to the inverse of their corresponding Fisher information because of the nature of PMLE. The derivation of these variance-covariance matrices is provided in Supp. Section S4.3.

The variance-covariance matrices of the stage-wise estimators can be estimated by the corresponding sandwich variance estimators. For example, the estimated variance-covariance matrix of $\widehat\bftheta_J$ is $\overline\bfI_{J,n}^{-1}\overline\bfV_{J,n}\overline\bfI_{J,n}^{-1}$, where $\overline\bfI_{J,n}= - n^{-1}\sum_{i=1}^n \ddot\ell_J(\bfthetahat_J;\calO_{J,i})$, and $\overline\bfV_{J,n}=n^{-1}\sum_{i=1}^n \dot \ell_J(\bfthetahat_J;\calO_{J,i})\dot \ell_J(\bfthetahat_J;\calO_{J,i})^\top$ with $\dot \ell_J = \nabla_{\bftheta_J}\ell_J$. The estimated variance-covariance matrices for other parameters have the same sandwich form: $\overline\bfI_{(j,J),\bfOmega,\bfOmega,n}^{-1}\overline\bfV_{(j,J),\bfOmega,n}\overline\bfI_{(j,J),\bfOmega,\bfOmega,n}^{-1}$ for $\bfOmega_{j,J}$, $j\in [J-1]$, and $\overline\bfI_{e_k,\bfgamma,\bfgamma,n}^{-1}\overline\bfV_{e_k,\bfgamma,n}\overline\bfI_{e_k,\bfgamma,\bfgamma,n}^{-1}$ for $\bfgamma_{e_k}$, $e_k\in \scrT_k$ with $k=2,\cdots,J-1$. Their expressions are given in Supp. Section S4.3.

\section{Simulation studies} \label{sec:sim}

{We evaluate the proposed vine copula-based framework via three simulation studies with $J=3$ event times. The first two studies assess the finite-sample performance of the stage-wise pseudo-maximum likelihood estimation (PMLE) procedure under heterogeneous copula families and covariate-dependent pairwise associations. Simulation Study I demonstrates the consistency of the proposed estimators and the validity of the analytic variance estimators under correct copula specification. Simulation Study II evaluates the robustness of the PMLE framework to copula misspecification. Simulation Study III compares the proposed framework with the multivariate Archimedean copula approach of \citet{Li2020}, where both models are correctly specified. Simulation Study III is provided in Supp. Section S5.3.}

\subsection{Simulation Study I}

{We generate two covariates $\bfZ_i=(Z_{1,i},Z_{2,i})^\top$ and three event times $(T_{1,i},T_{2,i},T_{3,i})^\top$ from a data-generating mechanism ensuring that (i) the marginal distribution of each $T_{j,i}$ given $\bfZ_i$ follows a Cox PH model, and (ii) pairwise joint distributions follow three different copula families: $\bbmC_{1,3}$ being Gumbel, $\bbmC_{2,3}$ being Clayton, and $\bbmC_{1,2|3}$ being Gaussian, all three with their parameters depending on covariates $\bfZ_i$. The Gumbel and Clayton copulas are Archimedean, whereas the Gaussian copula is elliptical; the design thus involves both Archimedean
and non-Archimedean families. These three copula families also represent distinct tail dependence structures: the Gumbel copula exhibits upper-tail dependence, the Clayton copula exhibits lower-tail dependence, and the Gaussian copula exhibits no tail dependence.
In addition, the Gaussian copula allows both positive and negative associations, whereas the Gumbel and Clayton copulas only model positive associations.}

{Specifically, $Z_{1,i}\sim\mathrm{Unif}(-0.5,0.5)$ and $Z_{2,i}\sim\mathrm{Bern}(1/3)$, a Bernoulli distribution with parameter $1/3$. Given $\bfZ_i$ , we generate event times $(T_{1,i},T_{2,i},T_{3,i})^\top$ following the steps below.}

\begin{description}
    \item [\bf Step 1: Generate $T_{3,i}$.] First, sample $\epsilon_{3,i}\sim \mathrm{Unif}(0,1)$. Then, obtain the event time $T_{3,i}$ using
    \[
        T_{3,i} = \exp\left[-\zeta_3-\bfbeta_3^\top\bfZ_i + \log\{-\log(1-\epsilon_{3,i})\}\right],
    \]
    where the parameters are $\zeta_3=-0.2$ and $\bfbeta_3=(2,2)^\top$. Next, calculate
    \[
        U_{3,\bfZ,i} = \exp\left(-e^{\zeta_3}T_{3,i} e^{\bfbeta_3^\top\bfZ_i}\right).
    \]
    \medskip

    \item [\bf Step 2: Generate $(U_{1|3,\bfZ,i}, U_{2|3,\bfZ,i})$ from a Gaussian copula.] The bivariate vector $(U_{1|3,\bfZ,i}, U_{2|3,\bfZ,i})$ is generated from a Gaussian copula with the covariate-dependent copula parameter given by
    \[
        \alpha_{(1,2|3),\bfZ,i} = \gamma_{(1,2|3),0} + \gamma_{(1,2|3),1}Z_{1,i} + \gamma_{(1,2|3),2}Z_{2,i},
    \]
    where $\bfgamma_{1,2|3} = (-0.24, 0.49, 0.49)^\top$.
    \medskip

\item[\bf Step 3: Generate $U_{j,\bfZ,i}$, $j=1,2$, from two Archimedean copulas.]

Let $\bbmC_{1,3}$ be Gumbel and $\bbmC_{2,3}$ be Clayton, with
covariate-dependent parameters
\[
\begin{aligned}
\alpha_{(1,3),\bfZ,i}
&=
\exp\left(
\gamma_{(1,3),0}
+\gamma_{(1,3),1}Z_{1,i}
+\gamma_{(1,3),2}Z_{2,i}
\right)+1,
&
\bfgamma_{1,3}&=(0.85,1,0.1)^\top,\\
\alpha_{(2,3),\bfZ,i}
&=
\exp\left(
\gamma_{(2,3),0}
+\gamma_{(2,3),1}Z_{1,i}
+\gamma_{(2,3),2}Z_{2,i}
\right),
&
\bfgamma_{2,3}&=(0.29,0.1,1)^\top.
\end{aligned}
\]
For each $j=1,2$, obtain $U_{j,\bfZ,i}$ by solving
\[
U_{j|3,\bfZ,i}
=
\dot{\bbmC}_{j,3}
\left(
u\mid U_{3,\bfZ,i};
\alpha_{(j,3),\bfZ,i}
\right)
\]
with respect to $u$, and set the solution equal to $U_{j,\bfZ,i}$.

\medskip

    \item [\bf Step 4: Generate $T_{j,i}$ from $U_{j,\bfZ,i}$.] For each $j \in \{1,2\}$, obtain the nonterminal event time using
    \[
        T_{j,i} = \exp\left[-\zeta_j - \bfbeta_j^\top\bfZ_i + \log\{-\log(U_{j,\bfZ,i})\}\right].
    \]
    The parameters are set to $\zeta_1=0.1$, $\zeta_2=0.4$, and $\bfbeta_1 = \bfbeta_2 = (2,2)^\top$.
\end{description}

{Under the above data-generating mechanism,  the strength of the pairwise dependence varies; by the well-established one-to-one correspondence between the copula parameter and Kendall's $\tau$, $\tau_{1,3}(\bfZ) \in (0.55, 0.85)$, $\tau_{2,3}(\bfZ)\in (0.35, 0.70)$, and $\tau_{1,2|3}(\bfZ) \in (-0.3,0.3)$.  These correspondences for all copula families considered in this article are provided in Supp. Section S1. In addition, we generate the administrative censoring time $A_i$ from a uniform distribution over $(1,6)$, which yields censoring rates of about 30\%, 25\%, and 10\% for $ T_1$, $ T_2$, and $ T_3$, respectively.}

We generate 500 data replications with sample sizes $n=500$, $1{,}000$, and $2{,}000$. {For each replication, we obtain the estimates of the model parameters and their analytic variance estimates under the correct specification of all copula families.} Table~\ref{tab:sim-1} reports the summary statistics over 500 estimates for the parameters $\{\bfbeta_1,\bfbeta_2,\bfbeta_3,\bfgamma_{1,3},\bfgamma_{2,3},\bfgamma_{1,2|3}\}$, and the summary statistics include {the empirical bias, standard deviation (SD), and average standard error (ASE).} In addition, we also report the empirical coverage proportion (CP) of the 95\% confidence interval (CI) using the proposed analytic variance estimator: $\widehat\theta\pm 1.96 \times \text{SE}_{\widehat\theta}$, where $\widehat\theta$ is the estimate of a parameter in $\bfOmega$, and $\text{SE}_{\widehat\theta}$ is the square root of its analytic variance estimator. Estimates for the baseline survival function $\exp\{-\Lambda_j(\cdot)\}$ are available in Web Table S1 in Supp. Section S5.1. {Supp. Section S5.1 also reports the computational time of the proposed estimation procedure.}

{Results in Table~\ref{tab:sim-1} and Web Table S1 show that the bias is small in absolute terms for all parameters, although $\gamma_{(1,3),2}$ and $\gamma_{(2,3),1}$, whose true value is $0.1$, exhibit a relatively larger proportional bias at $n=500$.  These biases diminish as the sample size increases, demonstrating the consistency of the estimation procedure.} In addition, the empirical SD and the ASE are close, and the empirical CP of the 95\% CIs remains at its nominal level, confirming the asymptotic normality and the validity of the analytic variance estimators.

\begin{table}[htbp]
    \centering
    \small
    \setlength{\tabcolsep}{3pt}
    {\begin{tabular}{l rrrc rrrc rrrc}
        \toprule
        & \multicolumn{4}{c}{$n=500$}& \multicolumn{4}{c}{$n=1{,}000$} & \multicolumn{4}{c}{$n=2{,}000$}  \\
        \cmidrule(lr){2-5} \cmidrule(lr){6-9} \cmidrule(lr){10-13}
        \multicolumn{1}{c}{Parameter} & \multicolumn{1}{c}{BIAS} & \multicolumn{1}{c}{SD} & \multicolumn{1}{c}{ASE} & CP & \multicolumn{1}{c}{BIAS} & \multicolumn{1}{c}{SD} & \multicolumn{1}{c}{ASE} & CP & \multicolumn{1}{c}{BIAS} & \multicolumn{1}{c}{SD} & \multicolumn{1}{c}{ASE} & CP \\
        \cmidrule(lr){1-1} \cmidrule(lr){2-5} \cmidrule(lr){6-9} \cmidrule(lr){10-13}
$\beta_{11}=2.00$ & 0.004 & 0.189 & 0.178 & 94.0 & 0.006 & 0.130 & 0.126 & 94.0 & 0.008 & 0.091 & 0.089 & 95.4 \\
$\beta_{12}=2.00$ & 0.005 & 0.127 & 0.122 & 94.6 & 0.004 & 0.088 & 0.086 & 94.6 & 0.001 & 0.061 & 0.061 & 95.0 \\
        \cmidrule(lr){1-1} \cmidrule(lr){2-5} \cmidrule(lr){6-9} \cmidrule(lr){10-13}
$\beta_{21}=2.00$ & 0.003 & 0.189 & 0.178 & 93.2 & 0.003 & 0.123 & 0.125 & 95.6 & 0.008 & 0.090 & 0.089 & 94.8 \\
$\beta_{22}=2.00$ & 0.005 & 0.131 & 0.121 & 94.2 & 0.008 & 0.089 & 0.086 & 93.4 & $-$0.003 & 0.060 & 0.061 & 95.6 \\
        \cmidrule(lr){1-1} \cmidrule(lr){2-5} \cmidrule(lr){6-9} \cmidrule(lr){10-13}
$\beta_{31}=2.00$ & 0.005 & 0.193 & 0.178 & 93.2 & 0.006 & 0.126 & 0.126 & 94.2 & 0.007 & 0.093 & 0.089 & 94.0 \\
$\beta_{32}=2.00$ & 0.010 & 0.129 & 0.121 & 94.0 & 0.008 & 0.089 & 0.086 & 93.0 & $-$0.001 & 0.059 & 0.061 & 95.4 \\
        \cmidrule(lr){1-1} \cmidrule(lr){2-5} \cmidrule(lr){6-9} \cmidrule(lr){10-13}
$\gamma_{(1,3),}=0.85$ & 0.005 & 0.094 & 0.096 & 94.4 & 0.003 & 0.064 & 0.067 & 96.2 & 0.004 & 0.046 & 0.047 & 94.6 \\
$\gamma_{(1,3),1}=1.00$ & 0.028 & 0.253 & 0.242 & 95.6 & 0.006 & 0.167 & 0.166 & 93.8 & 0.014 & 0.115 & 0.115 & 95.0 \\
$\gamma_{(1,3),2}=0.10$ & 0.029 & 0.155 & 0.151 & 93.4 & 0.020 & 0.101 & 0.103 & 96.0 & 0.010 & 0.072 & 0.072 & 94.8 \\
        \cmidrule(lr){1-1} \cmidrule(lr){2-5} \cmidrule(lr){6-9} \cmidrule(lr){10-13}
$\gamma_{(2,3),}=0.29$ & 0.008 & 0.138 & 0.144 & 94.6 & $-$0.005 & 0.099 & 0.102 & 95.4 & $-$0.003 & 0.071 & 0.072 & 95.4 \\
$\gamma_{(2,3),1}=0.10$ & 0.028 & 0.319 & 0.337 & 96.6 & 0.010 & 0.224 & 0.230 & 94.6 & $-$0.008 & 0.159 & 0.159 & 94.2 \\
$\gamma_{(2,3),2}=1.00$ & 0.016 & 0.192 & 0.200 & 95.2 & 0.007 & 0.133 & 0.138 & 95.6 & 0.008 & 0.097 & 0.097 & 94.8 \\
        \cmidrule(lr){1-1} \cmidrule(lr){2-5} \cmidrule(lr){6-9} \cmidrule(lr){10-13}
$\gamma_{(1,2| 3),}=-0.24$ & 0.005 & 0.063 & 0.062 & 94.4 & 0.002 & 0.045 & 0.044 & 95.6 & 0.000 & 0.031 & 0.031 & 95.0 \\
$\gamma_{(1,2| 3),1}=0.49$ & 0.002 & 0.183 & 0.170 & 93.4 & $-$0.007 & 0.122 & 0.121 & 95.2 & 0.002 & 0.088 & 0.086 & 94.8 \\
$\gamma_{(1,2| 3),2}=0.49$ & -0.006 & 0.099 & 0.102 & 96.2 & $-$0.002 & 0.074 & 0.072 & 94.6 & $-$0.000 & 0.053 & 0.052 & 94.2 \\
        \bottomrule
    \end{tabular}
    \caption{{Results of Simulation Study I -- summary statistics over 500 estimates. BIAS: empirical bias; SD:  empirical standard deviation; ASE: average standard error; CP: empirical coverage proportion of the 95\% CI, reported as a percentage.}} \label{tab:sim-1}
    }
\end{table}

\subsection{{Simulation Study II}}\label{subsec:simu2}

{In this study, we generate data with sample sizes of $500$ and $1{,}000$ using the same mechanism as in Simulation Study I but estimate parameters under misspecified copula families. We consider three copula-misspecification scenarios.}
\begin{description}
\item [{\bf Gu-Cl-Fr.}] {$\{\bbmC_{1,3},\bbmC_{2,3}, \bbmC_{1,2|3}\}=\{\mathrm{Gumbel},\mathrm{Clayton},\mathrm{Frank}\}$. Only the conditional between-nonterminal copula is misspecified to be Frank, which, similar to Gaussian, also exhibits no tail dependence and allows both positive and negative association. }

\item [{\bf Cl-Cl-Ga.}] {$\{\bbmC_{1,3},\bbmC_{2,3}, \bbmC_{1,2|3}\}=\{\mathrm{Clayton},\mathrm{Clayton},\mathrm{Gaussian}\}$. One nonterminal-terminal copula is misspecified. }

\item [{\bf Cl-Gu-Ga.}] {$\{\bbmC_{1,3},\bbmC_{2,3}, \bbmC_{1,2|3}\}=\{\mathrm{Clayton},\mathrm{Gumbel},\mathrm{Gaussian}\}$.  Both nonterminal-terminal copulas are misspecified.}
\end{description}

{Web Table S2 shows which copulas are correctly specified or misspecified under each scenario. We examine estimation bias and variability for both the marginal and dependence parameters. Table~\ref{tab:sim2} presents empirical bias and SD of the marginal regression coefficient estimates under correct specification (from Study I) and three copula misspecification scenarios. Several patterns emerge. First, the terminal marginal estimates ($\widehat\bfbeta_3$) remain unaffected across all misspecification scenarios, with results identical to those under correct specification. This robustness is expected: in the proposed stage-wise estimation, the terminal-event marginal is estimated directly from its own observed-data likelihood that does not involve any copula and is therefore inherently robust to copula misspecification. Second, for the nonterminal event times, bias arises selectively according to which nonterminal-terminal copula is misspecified. Specifically, the $T_1$ marginal estimates ($\widehat\beta_{11}$, $\widehat\beta_{12}$) show noticeable bias only in the two scenarios where $\bbmC_{1,3}$ is misspecified (Cl-Cl-Ga, Cl-Gu-Ga), while the $T_2$ marginal estimates ($\widehat\beta_{21}$, $\widehat\beta_{22}$) are affected only under Cl-Gu-Ga, the sole scenario in which $\bbmC_{2,3}$ is misspecified. This selective pattern directly reflects the separation structure of the stage-wise estimation procedure. At Stage 1, the marginal parameters for each nonterminal event are estimated jointly with the corresponding nonterminal-terminal copula only, so its estimation is unaffected by misspecification of the other copulas.}

{Third, all marginal estimates are fully robust to the misspecification of the between-nonterminal copula $\bbmC_{1,2|3}$ (Gu-Cl-Fr scenario), consistent with $\bbmC_{1,2|3}$ entering only at Stage 2 and having no influence on Stage 1 estimation. Finally, even in scenarios where misspecification induces bias, its magnitude remains small relative to the true coefficient value (e.g., bias $\approx -0.013$ to $-0.020$ versus true value of $2.00$). Importantly, the bias does not diminish as the sample size increases from $n = 500$ to $n = 1{,}000$ --- in fact, it tends to grow slightly in absolute magnitude (e.g., the magnitude of the bias of $\widehat\beta_{11}$ increases from $0.013$ to $0.020$, and the magnitude of the bias of $\widehat\beta_{22}$ from $0.013$ to $0.021$) --- consistent with misspecification bias being a structural rather than finite-sample phenomenon. By contrast, the empirical SD decreases with $n$ as expected, so the increasing bias is not offset by reduced variability, underscoring the importance of correctly specifying the nonterminal-terminal copula families for valid marginal inference.}

{For dependence parameters, the regression coefficients $\bfgamma$ are not directly comparable across copula families, as their interpretation depends on the assumed link function and copula family. Instead, we considered Kendall's $\tau$ implied by the fitted copula family via the one-to-one correspondence between the copula parameter and Kendall's $\tau$: $\tau = \tau(\alpha)$. Since the copula parameter is covariate-dependent, so is Kendall's tau, $\tau(\bfZ) = \tau\{g(\bfgamma^\top \bfZ)\}$, which is estimated by a plug-in estimator  $\widehat\tau(\bfZ) = \tau\{g(\widehat\bfgamma^\top \bfZ)\}$. We estimated all Kendall's $\tau$, $\tau_{1,3}$, $\tau_{2,3}$, and $\tau_{1,2|3}$ for each vine copula edge, at $Z_2 =0, 1$ and values of $Z_1$ ranging from $-$0.5 to 0.5 with increments 0.01, and calculated their empirical bias and SD over 500 replications. Figure~\ref{fig:simu2_tau_C12} and Web Figures S2 - S4 display the bias in the estimated covariate-dependent Kendall's $\tau$ under correct and incorrect copula specifications.}

\begin{table}[htbp]
\centering
{\begin{tabular}{lcccc}
\toprule
Parameter & \multicolumn{1}{c}{Correct} & \multicolumn{1}{c}{Gu-Cl-Fr} & \multicolumn{1}{c}{Cl-Cl-Ga} & \multicolumn{1}{c}{Cl-Gu-Ga} \\
\hline
\multicolumn{5}{c}{$n = 500$} \\
\multicolumn{5}{c}{$T_1$ marginal model} \\
\quad $\beta_{11} = 2.00$ & 0.004 (0.189) & 0.004 (0.189) & {\bf $-$0.013 (0.203)} & {\bf $-$0.013 (0.203)} \\
\quad $\beta_{12} = 2.00$ & 0.005 (0.127) & 0.005 (0.127) & {\bf $-$0.005 (0.141)} & {\bf $-$0.005 (0.141)} \\
\multicolumn{5}{c}{$T_2$ marginal model} \\
\quad $\beta_{21} = 2.00$ & 0.003 (0.189) & 0.003 (0.189) & 0.003 (0.189) & {\bf $-$0.011 (0.186)}\\
\quad $\beta_{22} = 2.00$ & 0.005 (0.131) & 0.005 (0.131) & 0.005 (0.131) & {\bf 0.013 (0.124)} \\
\multicolumn{5}{c}{$T_3$ marginal model} \\
\quad $\beta_{31} = 2.00$ & 0.005 (0.193) & 0.005 (0.193) & 0.005 (0.193) & 0.005 (0.193) \\
\quad $\beta_{32} = 2.00$ & 0.010 (0.129) & 0.010 (0.129) & 0.010 (0.129) & 0.010 (0.129)  \\
\hline
\multicolumn{5}{c}{$n = 1000$} \\
\multicolumn{5}{c}{$T_1$ marginal model} \\
\quad $\beta_{11} = 2.00$ & 0.006 (0.130) & 0.006 (0.130) & {\bf $-$0.020 (0.141) } & {\bf $-$0.020 (0.141)} \\
\quad $\beta_{12} = 2.00$ & 0.004 (0.088) & 0.004 (0.088) & {\bf $-$0.014 (0.097)} & {\bf $-$0.014 (0.097)}  \\
\multicolumn{5}{c}{$T_2$ marginal model} \\
\quad $\beta_{21} = 2.00$ & 0.003 (0.123) & 0.003 (0.123) & 0.003 (0.123) & {\bf $-$0.009 (0.124)} \\
\quad $\beta_{22} = 2.00$ & 0.008 (0.089) & 0.008 (0.089) & 0.008 (0.089) & {\bf 0.021 (0.089)} \\
\multicolumn{5}{c}{$T_3$ marginal model} \\
\quad $\beta_{31} = 2.00$ & 0.006 (0.126) & 0.006 (0.126) & 0.006 (0.126) & 0.006 (0.126)  \\
\quad $\beta_{32} = 2.00$ & 0.008 (0.089) & 0.008 (0.089) & 0.008 (0.089) & 0.008 (0.089) \\
\bottomrule
\end{tabular}}

\caption{{Results of Simulation Study II -- empirical bias and SD (shown in parentheses) over $500$ estimates of marginal regression coefficients under correct and incorrect copula specifications. The bolded numbers are the summary statistics from the incorrect specification of the copula family used to model the joint distribution of the corresponding nonterminal and terminal event times.}}
\label{tab:sim2}
\end{table}

{For the nonterminal-terminal dependence parameters (Web Figures S2 and S3), the bias pattern mirrors that of the marginal coefficients: the bias in $\widehat{\tau}_{1,3}$ is noticeable only in scenarios where $\bbmC_{1,3}$ is misspecified (Cl-Cl-Ga, Cl-Gu-Ga), while $\widehat{\tau}_{2,3}$ is affected only under Cl-Gu-Ga, the sole scenario in which $\bbmC_{2,3}$ is misspecified. Each nonterminal-terminal $\tau$ estimate is robust to the misspecification of all other copulas in the vine, a direct consequence of the first-tree copulas being estimated separately at Stage 1. Furthermore, the bias in $\widehat{\tau}_{1,3}$ and $\widehat{\tau}_{2,3}$ varies with $Z_1$ and $Z_2$. At certain covariate values, misspecification leads to more severe, non-ignorable bias relative to the estimation variability, as evidenced by the $\pm 1.96\times \text{SD}$ ribbon failing to cover zero in Web Figures S2 and S3.  As with the marginal parameters, the bias does not diminish as $n$ increases from $500$ to $1{,}000$, confirming that it reflects a structural consequence of copula misspecification rather than a finite-sample artifact.}

\begin{figure}[htbp]
\centering
\includegraphics[width = 1\textwidth]{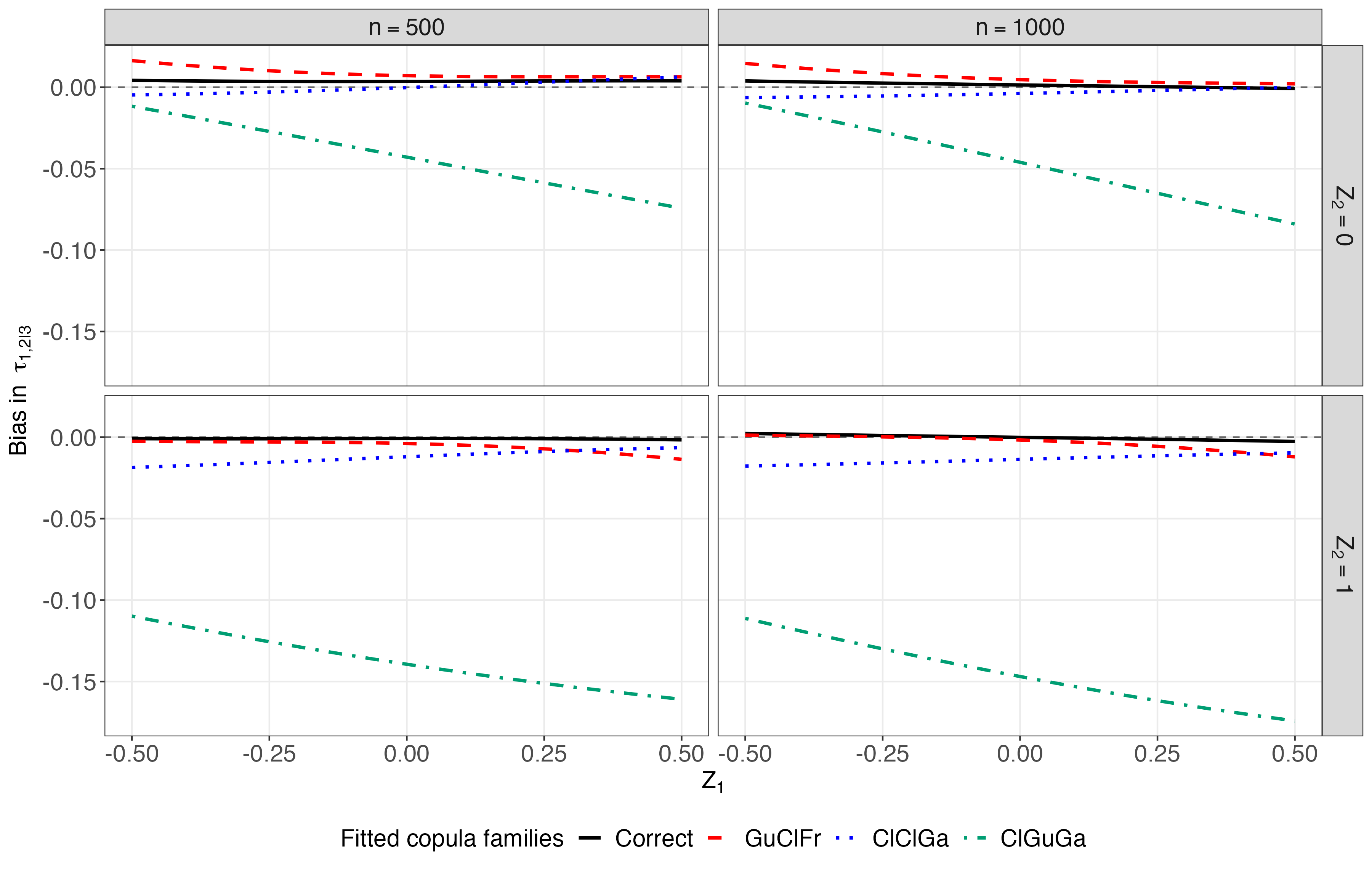}
\caption{{Results of Simulation Study II -- empirical bias over 500 estimates of covariate-dependent $\tau_{1,2|3}$, between-nonterminal Kendall's $\tau$ given $T_3$ under correct and incorrect copula specifications.}}\label{fig:simu2_tau_C12}
\end{figure}

{For the conditional between-nonterminal dependence $\tau_{1,2|3}$ (Figure~\ref{fig:simu2_tau_C12} and Web Figure S4), the bias results from two sources. The first is due to direct misspecification of $\bbmC_{1,2|3}$ alone under the Gu-Cl-Fr scenario. In this scenario, the bias is not substantial because the Frank and Gaussian families are both symmetric with no tail dependence, representing a mild degree of misspecification. The second source is propagated: even when $\bbmC_{1,2|3}$ is correctly specified, misspecification of a first-tree copula ($\bbmC_{1,3}$ or $\bbmC_{2,3}$), e.g., Cl-Cl-Ga and Cl-Gu-Ga scenarios, biases the conditional survival functions $\widehat{U}_{j|3,\bfZ}=\dot\bbmC_{j,3}(\widehat{U}_{j,\bfZ}|\widehat{U}_{3,\bfZ};\widehat\alpha_{j,3,\bfZ})$, $j=1,2$, that serve as pseudo-observations for Stage 2 estimation, carrying the error forward to $\widehat{\tau}_{1,2|3}$.  Furthermore, the propagated effects accumulate: the bias under Cl-Gu-Ga, where both $\bbmC_{1,3}$ and $\bbmC_{2,3}$ are misspecified, is larger than that under Cl-Cl-Ga, where only $\bbmC_{1,3}$ is misspecified.}

\section{A real data example} \label{sec:data}

We apply the proposed vine-copula-based modeling framework to analyze data from Indiegogo, one of the largest crowdfunding platforms \citep{Zhang2012}. In a crowdfunding campaign, the creator sets a fundraising goal and corresponding rewards or perks, and interacts with backers during the fundraising period \citep{Su2023}. These interactions include updates from the creator on the latest campaign progress \citep{Herd2022} and discussions between the creator and backers through comments and responses \citep{Dai2019, Cornelius2020, Fan2020}. Thus, in this analysis, we study three essential types of creator-backer interactions that were extensively studied as nonterminal events: (i) the first update of the campaign \citep{Herd2022}, (ii) the first 10 comments from either backers or the creator \citep{Fan2020}, and (iii) the first response from a creator to a comment \citep{Cornelius2020}. The terminal event is the creator's departure from the platform.

The relationships among these nonterminal events, as well as their connection to the terminal event, are also widely explored. For example, \citet{Camacho2019} found that when the creator updates more, the campaign also receives more comments, and \citet{Jabr2022} proposed that excessive information may lead to information overload in crowdfunding campaigns. Therefore, understanding these dynamics can help creators better manage their relationships and interactions with backers, ultimately maximizing the amount raised. The existing literature suggests that creators' experience in their first campaign has a significant impact on their future campaigns \citep{Kim2020}. Thus, we focus on creators with multiple sequential campaigns and set the end of their first campaign as the baseline. Let $T_\rmU$, $T_{10\rmC}$, $T_\rmR$, and $T_\rmD$ denote the time from the baseline to the first update, to the first 10 comments, to the first response to comments, and departure, respectively. We also selected several covariates related to a creator's first campaign that may influence the nonterminal and terminal events: (i) the length of the narrative/campaign description \citep{Wei2022}, (ii) the amount raised \citep{Liu2025}, (iii) the number of updates \citep{Herd2022}, (iv) the mean perk price \citep{Su2023},\footnote{An Indiegogo (crowdfunding) campaign usually offers multiple perks. Each perk is associated with a specific price.} (v) the number of comments \citep{Gao2025}, (vi) whether the creator responds to comments \citep{Kozinets2010}, and (vii) whether the platform verifies the creator's email address \citep{Beck2024}.

\subsection{Data description}

The data consist of $6{,}303$ creators that (i) entered the platform between 2012 and 2014, (ii) had more than one campaign, and (iii) did not have missing values for the above covariates. The departure time is fully observed for all creators, and the censoring rates for $T_\rmU$, $T_{10\rmC}$, and $T_\rmR$ are 50\%, 81\%, and 91\%, respectively. Two binary covariates are coded: whether a creator's email address is verified (47\%) and whether a creator responded during the first campaign (1\%). The other five covariates are continuous and heavily right-skewed; therefore, we used the log-transformed values in the analysis. Web Table S4 in Supp. Section S6.1 reports the summary statistics of the five log-transformed variables.

\subsection{Bivariate nonterminal event times} \label{subsec:bivariate}

We begin with the analysis of three pairs of bivariate nonterminal event times. We constructed a vine graph for each pair: $\scrV_{\rmU,10\rmC}$ for $(T_{\rmU},T_{10\rmC},T_{\rmD})$, $\scrV_{\rmR,10\rmC}$ for $(T_{\rmR},T_{10\rmC},T_{\rmD})$, and $\scrV_{\rmU,\rmR}$ for $(T_{\rmU},T_{\rmR},T_{\rmD})$, outlined in Web Figure S5 of Supp. Section S6.1. For each bivariate copula, we consider four candidate families: Clayton, Gumbel, Frank, and Gaussian. The family with the maximum log-likelihood evaluated at the corresponding estimate is used in the analysis (reported in Web Table S5 in Supp. Section S6.1). The selected families for $\bbmC_{\rmU,\rmD}$, $\bbmC_{\rmR,\rmD}$, and $\bbmC_{10\rmC,\rmD}$ are all Gumbel; $\bbmC_{\rmU,10\rmC|\rmD}$ is Gumbel while $\bbmC_{\rmR,10\rmC|\rmD}$ and $\bbmC_{\rmU,\rmR|\rmD}$ are Clayton.

Tables~\ref{tab:marginal} and~\ref{tab:copula} report the estimated covariate effects on marginals and bivariate associations, respectively. The standard errors of the estimates and two-sided $p$-values are also provided. Here, we highlight some observations that align with the existing literature and reserve the rest for Supp. Section S6.2. The email verification status has a significant negative effect on all four marginals. \citet{Beck2024} suggest that identity verification (e.g., email verification for campaign creators) can help build trust among backers. Therefore, when creators use email verification to build backers' trust, they may rely less on updates, responses, and comments to achieve the same goal. Also, since the creators have made extra efforts beyond a specific campaign to complete the email verification, they intend to manage the platform well to build a long-term relationship rather than leave (depart) early \citep{Palmatier2009}. The email verification status, on the other hand, has a significant negative effect on $\bbmC_{\rmU,\rmD}$, $\bbmC_{\rmR,\rmD}$, and $\bbmC_{10\rmC,\rmD}$. This could be explained by the dependence of a creator's lifetime on the trust establishment with backers, which can be achieved through two major sources. The first is achieved through direct interactions, such as updates, responses, and comments \citep{Acar2021}, while the second is facilitated by identity verification, including email verification \citep{Beck2024}. When creators rely more on email verification to build backers' trust, they will put less effort into direct interactions that foster trust. Consequently, these activities are less likely to be factors associated with creators' departure.

Web Figure S6 in Supp. Section S6.1 plots the estimated survival curves of $T_\rmU$, $T_\rmR$, $T_{10\rmC}$, and $T_{\rmD}$, which are functions of the covariates. We select four ``representative" groups of creators: all four groups have median values for five continuous covariates, but they vary by their email verification and response statuses during the first campaign. Based on the above results, we conducted additional investigations in Supp. Sections S6.3 and S6.4, which include the following. {First, we extend Kendall's $\tau$ to quantify the association between each pair of nonterminal events conditional on the terminal event time and investigate how this association varies across survival strata of creators with different lifetimes.} Second, we develop various models to predict the risk of departure based on its association with nonterminal events, modeled by the vine copula. In addition, we evaluate and compare their prediction performance using the area under the receiver operating characteristic curve.

\begin{table}[htbp]
    \centering
    \begin{tabular}{r rrr rrr}
    \toprule
    & \multicolumn{3}{c}{First Update} & \multicolumn{3}{c}{First Response} \\
    \cmidrule(lr){2-4} \cmidrule(lr){5-7}
    \multicolumn{1}{c}{Covariates} & \multicolumn{1}{c}{Est.} & \multicolumn{1}{c}{SE} & \multicolumn{1}{c}{$p$-value} & \multicolumn{1}{c}{Est.} & \multicolumn{1}{c}{SE} & \multicolumn{1}{c}{$p$-value} \\
    \cmidrule(lr){1-1} \cmidrule(lr){2-4} \cmidrule(lr){5-7}
    Email veri.: Yes vs No & $-$0.448 & 0.031 & \textbf{$<$0.001} & $-$0.488 & 0.117 & \textbf{$<$0.001} \\
    Response: Yes vs No &  0.189 & 0.166 & 0.255 &  0.532 & 0.238 & \textbf{0.025} \\
    Amount raised &  0.042 & 0.012 & \textbf{$<$0.001} &  0.166 & 0.038 & \textbf{$<$0.001} \\
    Narrative length &  0.015 & 0.016 & 0.329 &  0.172 & 0.043 & \textbf{$<$0.001} \\
    Number of updates &  0.038 & 0.006 & \textbf{$<$0.001} & $-$0.010 & 0.014 & 0.476\\
    Mean perk price & $-$0.026 & 0.012 & \textbf{0.035} &  0.005 & 0.021 & 0.801\\
    Number of comments & $-$0.033 & 0.006 & \textbf{$<$0.001} & $-$0.058 & 0.014 & \textbf{$<$0.001} \\
    \cmidrule(lr){1-1} \cmidrule(lr){2-4} \cmidrule(lr){5-7}
    & \multicolumn{3}{c}{First 10 Comments} &
    \multicolumn{3}{c}{Departure}\\
    \cmidrule(lr){1-1} \cmidrule(lr){2-4} \cmidrule(lr){5-7}
    Email veri.: Yes vs No & $-$0.337 & 0.052 & \textbf{$<$0.001} & $-$0.580 & 0.027 & \textbf{$<$0.001} \\
    Response: Yes vs No &  0.120 & 0.204 & 0.556 &  0.124 & 0.158 & 0.434\\
    Amount raised &  0.090 & 0.026 & \textbf{$<$0.001} &  0.065 & 0.011 & \textbf{$<$0.001} \\
    Narrative length &  0.018 & 0.042 & 0.673 & $-$0.003 & 0.011 & 0.765\\
    Number of updates &  0.073 & 0.013 & \textbf{$<$0.001} &  0.005 & 0.004 & 0.219\\
    Mean perk price & $-$0.030 & 0.021 & 0.161 & $-$0.027 & 0.010 & \textbf{0.008} \\
    Number of comments &  0.120 & 0.023 & \textbf{$<$0.001} & $-$0.029 & 0.005 & \textbf{$<$0.001} \\
    \bottomrule
    \end{tabular}
    \caption{Covariate effects on marginal survival functions. Bold fonts indicate statistical significance at the level of 5\%.} \label{tab:marginal}
\end{table}

\begin{table}[htbp]
    \centering
    \small
    \setlength{\tabcolsep}{3pt}
    \begin{tabular}{r rrr rrr rrr}
    \toprule
    & \multicolumn{3}{c}{$\bbmC_{\mathrm{U,D}}$: Gumbel} & \multicolumn{3}{c}{$\bbmC_{\mathrm{R,D}}$: Gumbel} & \multicolumn{3}{c}{$\bbmC_{\mathrm{10C,D}}$: Gumbel} \\
    \cmidrule(lr){2-4} \cmidrule(lr){5-7} \cmidrule(lr){8-10}
    \multicolumn{1}{c}{Covariates} & \multicolumn{1}{c}{Est.} & \multicolumn{1}{c}{SE} & \multicolumn{1}{c}{$p$-value} & \multicolumn{1}{c}{Est.} & \multicolumn{1}{c}{SE} & \multicolumn{1}{c}{$p$-value} & \multicolumn{1}{c}{Est.} & \multicolumn{1}{c}{SE} & \multicolumn{1}{c}{$p$-value} \\
    \cmidrule(lr){1-1} \cmidrule(lr){2-4} \cmidrule(lr){5-7} \cmidrule(lr){8-10}
    Email veri.: Yes vs No & $-$0.761 & 0.095 & \textbf{$<$0.001} & $-$0.467 & 0.210 & \textbf{0.026} & $-$0.819 & 0.136 & \textbf{$<$0.001} \\
    Response: Yes vs No &  0.833 & 0.370 & \textbf{0.025} & $-$0.067 & 0.305 & 0.827 &  1.166 & 0.380 & \textbf{0.002} \\
    Amount raised & $-$0.022 & 0.038 & 0.549 &  0.018 & 0.062 & 0.767 &  0.010 & 0.055 & 0.850 \\
    Narrative length &  0.153 & 0.089 & 0.087 &  0.165 & 0.051 & \textbf{0.001} & $-$0.009 & 0.094 & 0.920\\
    Number of updates &  0.008 & 0.018 & 0.632 & $-$0.076 & 0.028 & \textbf{0.007} &  0.040 & 0.024 & 0.100\\
    Mean perk price &  0.046 & 0.034 & 0.176 &  0.100 & 0.057 & 0.080 &  0.026 & 0.044 & 0.552\\
    Number of comments & $-$0.024 & 0.020 & 0.227 & $-$0.011 & 0.032 & 0.738 &  0.054 & 0.026 & \textbf{0.041} \\
    \cmidrule(lr){1-1} \cmidrule(lr){2-4} \cmidrule(lr){5-7} \cmidrule(lr){8-10}
    & \multicolumn{3}{c}{$\bbmC_{\mathrm{U,10C}|\mathrm{D}}$: Gumbel} & \multicolumn{3}{c}{$\bbmC_{\mathrm{R,10C}|\mathrm{D}}$: Clayton} & \multicolumn{3}{c}{$\bbmC_{\mathrm{U,R}|\mathrm{D}}$: Clayton} \\
    \cmidrule(lr){1-1} \cmidrule(lr){2-4} \cmidrule(lr){5-7} \cmidrule(lr){8-10}
    Email veri.: Yes vs No&  0.240 & 0.186 & 0.197 & $-$0.020 & 0.205 & 0.923 & $-$0.001 & 0.642 & 0.999\\
    Response: Yes vs No& $-$0.677 & 0.560 & 0.227 &  0.657 & 0.491 & 0.181 & $-$0.480 & 0.875 & 0.583\\
    Amount raised&  0.031 & 0.087 & 0.724 & $-$0.129 & 0.063 & \textbf{0.039} & $-$0.313 & 0.206 & 0.129\\
    Narrative length&  0.010 & 0.069 & 0.886 &  0.070 & 0.078 & 0.375 &  1.041 & 0.410 & \textbf{0.011} \\
    Number of updates& $-$0.016 & 0.030 & 0.606 & $-$0.064 & 0.030 & \textbf{0.035} & $-$0.306 & 0.118 & \textbf{0.010} \\
    Mean perk price& $-$0.054 & 0.053 & 0.307 & $-$0.026 & 0.077 & 0.733 & $-$0.208 & 0.164 & 0.206\\
    Number of comments&  0.122 & 0.051 & \textbf{0.018} & $-$0.033 & 0.028 & 0.243 &  0.866 & 0.311 & \textbf{0.005} \\
    \bottomrule
    \end{tabular}
    \caption{Covariate effects on bivariate copulas. R: first response; U: first update; 10C: first 10 comments; D: departure. Bold fonts indicate statistical significance at the level of 5\%.} \label{tab:copula}
\end{table}

\subsection{Trivariate nonterminal event times}

We move on to the analysis of all three nonterminal event times and are particularly interested in the association between $T_{\rmU}$ and $T_{\rmR}$ given $T_{10\rmC}$ and $T_{\rmD}$. Creators interact with backers through campaign updates or responses to individual comments. However, when creators use both ways simultaneously, it may cause information overload and harm the campaigns \citep{Jabr2022}. Thus, we build the vine copula graph shown in Web Figure S1 in Supp. Section S3 with $4=\rmD$, $3=10\rmC$, $2=\rmR$, and $1=\rmU$, which allows us to study $\bbmC_{\rmU,\rmR|10\rmC,\rmD}$ in the last tree. In Section~\ref{subsec:bivariate}, we have estimated all the marginals and bivariate copulas of the first two trees: $\{\bbmC_{\rmU,\rmD},\bbmC_{\rmR,\rmD},\bbmC_{10\rmC,\rmD},\bbmC_{\rmU,10\rmC|\rmD},\bbmC_{\rmR,10\rmC|\rmD}\}$, reported in Tables~\ref{tab:marginal} and~\ref{tab:copula}. Thus, here, we estimate the parameters associated with the bivariate copula $\bbmC_{\rmU, \rmR|10\rmC,\rmD}$.

Web Table S6 in Supp. Section S6.1 reports the estimates of covariate effects under each of the four copula families: Clayton, Gumbel, Frank, and Gaussian. These estimates show that none of the covariate effects are significant under any copula family. Thus, we assume that $\alpha_{\rmU, \rmR|10\rmC,\rmD,\bfZ}=\alpha_{\rmU, \rmR|10\rmC,\rmD}$ is a constant. Under this assumption, we select Frank as the optimal copula family for $\bbmC_{\rmU, \rmR|10\rmC,\rmD}$ because it has the largest log-likelihood among the four families; their log-likelihood values are reported in Web Table S5. {
Under Frank, the estimate of $\alpha_{\rmU,\rmR|10\rmC,\rmD}$ is $-0.303$ with a standard error of $0.200$ and a two-sided $p$-value of $0.13$ for testing $H_0: \alpha_{\rmU,\rmR|10\rmC,\rmD} = 0$. While this provides weak evidence of a negative association—resonating with the findings of \citet{Jabr2022} that creators avoid information redundancy by substituting updates for responses—this non-significance at the level of $0.05$ could be attributed to a power deficiency. Because estimating $\bbmC_{\rmU,\rmR|10\rmC,\rmD}$ in the final tree requires subjects to have observed trajectories for all prerequisite events, the extreme censoring rates of the nonterminal events ($50\%$, $81\%$, and $91\%$) restrict the effective sample size to $231$ out of $6{,}303$ creators. Consequently, the test is underpowered, and the non-significant $p$-value should be interpreted with caution because severe information loss substantially limits the ability to detect a nonzero conditional association.
}

\section{Discussion} \label{sec:disc}

We develop a flexible modeling framework for {analyzing multivariate event times under informative censoring} based on the vine copula, where the multivariate dependence among $\bfT_{[J]}$ is decomposed into a collection of pairwise dependencies encoded by a vine graph with the root node of the first tree corresponding to the terminal event time $T_J$. {The proposed framework accommodates heterogeneous pairwise dependence by allowing each event pair to be modeled with its own copula family and dependence parameter, while enabling  regression modeling of all marginal survival distributions to estimate direct covariate effects. The accompanying stage-wise estimation procedure substantially reduces the computational burden relative to maximum likelihood estimation, which estimates all parameters simultaneously. Moreover, its separation structure provides robustness to localized copula misspecification: misspecification of one first-tree edge does not affect the terminal marginal, the other first-tree edges, or their marginals, though it does carry into higher-tree quantities through the pseudo-observations, as Study II quantifies. In addition, we derive a counting process representation of {the likelihood} and introduce a conditional Kendall's $\tau$ that varies across survival strata, thereby distinguishing short-lived and long-lived subjects. These results may be useful for future methodological developments in survival analysis.}

The current work has some limitations that can be improved in future research. {First, although this study proposes a modeling framework to capture the heterogeneity of dependencies, more flexible model specifications can be adopted. Specifically, bivariate copulas can be specified nonparametrically to avoid misspecification arising from parametric copula families. For more complex covariate effects on the marginals and copula parameters, one could consider sieve-based specifications, e.g., the bundled sieve approach for marginals developed in \citet{Zhao2017} and the sieve-within-parametric approach for copulas used in \citet{Zhang2010} and \citet{Li2023}.} The stage-wise estimation procedure remains valid for these model specifications; however, the $n^{1/2}$-convergence and analytical variance estimators are likely unattainable \citep{Li2023}. Second, the vine copula-based method requires numerical integrations to obtain the joint survival and log-likelihood functions because it relies on a closed-form expression for the joint density of $\bfT_{[J]}$. The numerical integration operation can be computationally intensive and unwieldy when $J$ and $n$ are large. { Furthermore, applying higher-order vine trees to heavily censored data inherently suffers from compounding information loss. Because higher-order copulas are evaluated conditionally on the occurrences of prior events, heavy censoring of the individual event times results in a multiplicative reduction in the effective sample size available to estimate the higher-tree dependence parameters. As demonstrated in our data application, this structural limitation severely inflates the standard errors of conditional copula parameters, potentially resulting in underpowered statistical tests for conditional associations in higher-order trees.}

Third, a model diagnostic tool for assessing the goodness-of-fit of bivariate parametric copulas can serve as a valuable complement to the proposed modeling framework. {Fourth, in our data example, the terminal event time was fully observed, which significantly reduces the estimation difficulty. In applications where the terminal event time is subject to administrative censoring, evaluating the pseudo-log-likelihood for higher-order trees requires numerical integration over the unobserved region of the terminal event. Because the terminal event acts as the root node of the first tree, its pseudo-observation $u_J$ is recursively embedded in all higher-order conditional copulas, making this numerical integration computationally intensive. It could be more cost-effective to bypass this integration by adopting multiple imputation approaches \citep{taylor2002Survival,hsu2006survival,hsu2009nonparametric} on the censored time to the terminal event.} Lastly, the copula parameters in the current model are allowed to be dependent on covariates to capture between-individual heterogeneity, which may also be expanded to be event-time dependent, thereby relaxing the simplifying assumption.

{Our vine copula framework and the multivariate and nested frameworks of \cite{Li2020} and \cite{Li2023} are complementary approaches to modeling dependence among nonterminal event times. Consequently, they naturally lead to different between-nonterminal dependence summaries: under the multivariate and nested copula frameworks, a natural summary is the marginal Kendall's $\tau$, whereas the vine copula framework gives rise to the conditional Kendall's $\tau$. These measures address different scientific questions: the marginal Kendall's $\tau$ summarizes the overall association between two nonterminal events in the population, while the conditional Kendall's $\tau$ quantifies their residual association after accounting for the terminal event. We further extend the conditional Kendall's $\tau$ by allowing this residual association to vary across survival strata, thereby distinguishing short-lived and long-lived subgroups. Together, these three dependence summaries provide complementary perspectives on between-nonterminal dependence, ranging from the overall population association, to the residual association after accounting for the terminal event, to the heterogeneity of the residual association across survival strata. A more systematic theoretical investigation of the relationships among these dependence summaries and their scientific interpretations remains an important direction for future research.
}

\section*{Supplementary Materials}

The Supplementary Materials are provided as an ancillary file with this article. Section S1 lists commonly used parametric copulas, Section S2 gives the counting process representation of the log-likelihood, Section S3 presents the vine graph with $J=4$, Section S4 proves Propositions 1--3, and Sections S5 and S6 report additional simulation and data analysis results.

\bibliographystyle{jasa3}
\bibliography{TMIC}

\begin{thebibliography}{54}
\newcommand{\enquote}[1]{``#1''}
\expandafter\ifx\csname natexlab\endcsname\relax\def\natexlab#1{#1}\fi
\expandafter\ifx\csname url\endcsname\relax
  \def\url#1{{\tt #1}}\fi
\expandafter\ifx\csname urlprefix\endcsname\relax\def\urlprefix{URL }\fi

\bibitem[\protect\citeauthoryear{Acar, Dahl, Fuchs, and Schreier}{Acar
  et~al.}{2021}]{Acar2021}
Acar, O.~A., Dahl, D.~W., Fuchs, C., and Schreier, M. (2021), \enquote{The
  signal value of crowdfunded products,} {\em Journal of Marketing Research\/},
  58, 644--661.

\bibitem[\protect\citeauthoryear{Andersen, Borgan, Gill, and Keiding}{Andersen
  et~al.}{1993}]{Andersen1993}
Andersen, P.~K., Borgan, {\O}., Gill, R.~D., and Keiding, N. (1993), {\em
  Statistical Models Based on Counting Processes\/}, Springer Series in
  Statistics, New York, NY: Springer New York.

\bibitem[\protect\citeauthoryear{Andersen and Gill}{Andersen and
  Gill}{1982}]{AndersenGill1982}
Andersen, P.~K. and Gill, R.~D. (1982), \enquote{Cox's Regression Model for
  Counting Processes: A Large Sample Study,} {\em The Annals of Statistics\/},
  10, 1100--1120.

\bibitem[\protect\citeauthoryear{Arachchige, Chen, and Zhou}{Arachchige
  et~al.}{2025}]{Arachchige2024}
Arachchige, S.~J., Chen, X., and Zhou, Q.~M. (2025), \enquote{Two-stage pseudo
  maximum likelihood estimation of semiparametric copula-based regression
  models for semi-competing risks data,} {\em Lifetime Data Analysis\/}, 31,
  52--75.

\bibitem[\protect\citeauthoryear{Barthel, Geerdens, Czado, and Janssen}{Barthel
  et~al.}{2019}]{Barthel2019}
Barthel, N., Geerdens, C., Czado, C., and Janssen, P. (2019),
  \enquote{{Dependence modeling for recurrent event times subject to
  right-censoring with D-vine copulas},} {\em Biometrics\/}, 75, 439--451.

\bibitem[\protect\citeauthoryear{Barthel, Geerdens, Killiches, Janssen, and
  Czado}{Barthel et~al.}{2018}]{Barthel2018}
Barthel, N., Geerdens, C., Killiches, M., Janssen, P., and Czado, C. (2018),
  \enquote{Vine copula based likelihood estimation of dependence patterns in
  multivariate event time data,} {\em Computational Statistics \& Data
  Analysis\/}, 117, 109--127.

\bibitem[\protect\citeauthoryear{Beck, Wuyts, and Jap}{Beck
  et~al.}{2024}]{Beck2024}
Beck, B.~B., Wuyts, S., and Jap, S. (2024), \enquote{Guardians of trust: How
  review platforms can fight fakery and build consumer trust,} {\em Journal of
  Marketing Research\/}, 61, 682--699.

\bibitem[\protect\citeauthoryear{Bedford and Cooke}{Bedford and
  Cooke}{2002}]{Bedford2002}
Bedford, T. and Cooke, R.~M. (2002), \enquote{{Vines--a new graphical model for
  dependent random variables},} {\em The Annals of Statistics\/}, 30,
  1031--1068.

\bibitem[\protect\citeauthoryear{Camacho, Nam, Kannan, and Stremersch}{Camacho
  et~al.}{2019}]{Camacho2019}
Camacho, N., Nam, H., Kannan, P., and Stremersch, S. (2019),
  \enquote{Tournaments to crowdsource innovation: The role of moderator
  feedback and participation intensity,} {\em Journal of Marketing\/}, 83,
  138--157.

\bibitem[\protect\citeauthoryear{Chen, Fan, Pouzo, and Ying}{Chen
  et~al.}{2010}]{Chen2010joe}
Chen, X., Fan, Y., Pouzo, D., and Ying, Z. (2010), \enquote{Estimation and
  model selection of semiparametric multivariate survival functions under
  general censorship,} {\em Journal of Econometrics\/}, 157, 129--142.
  Nonlinear and Nonparametric Methods in Econometrics.

\bibitem[\protect\citeauthoryear{Chen}{Chen}{2012}]{Chen2012}
Chen, Y.-H. (2012), \enquote{Maximum likelihood analysis of semicompeting risks
  data with semiparametric regression models,} {\em Lifetime Data Analysis\/},
  18, 36--57.

\bibitem[\protect\citeauthoryear{Cornelius and Gokpinar}{Cornelius and
  Gokpinar}{2020}]{Cornelius2020}
Cornelius, P.~B. and Gokpinar, B. (2020), \enquote{The role of customer
  investor involvement in crowdfunding success,} {\em Management Science\/},
  66, 452--472.

\bibitem[\protect\citeauthoryear{Cox}{Cox}{1972}]{cox1972regression}
Cox, D. (1972), \enquote{{Regression models and life-tables},} {\em Journal of
  the Royal Statistical Society. Series B (Methodological)\/}, 34, 187--220.

\bibitem[\protect\citeauthoryear{Czado and Nagler}{Czado and
  Nagler}{2022}]{Czado2022annualrev}
Czado, C. and Nagler, T. (2022), \enquote{Vine copula based modeling,} {\em
  Annual Review of Statistics and Its Application\/}, 9, 453--477.

\bibitem[\protect\citeauthoryear{Dai and Zhang}{Dai and Zhang}{2019}]{Dai2019}
Dai, H. and Zhang, D.~J. (2019), \enquote{Prosocial goal pursuit in
  crowdfunding: Evidence from Kickstarter,} {\em Journal of Marketing
  Research\/}, 56, 498--517.

\bibitem[\protect\citeauthoryear{Davis, Li, Wai, Tyldesley, Simmons, Baliski,
  and McBride}{Davis et~al.}{2014}]{Davis2014}
Davis, M.~K., Li, D., Wai, E., Tyldesley, S., Simmons, C., Baliski, C., and
  McBride, M.~L. (2014), \enquote{Hospital-related cardiac morbidity among
  survivors of breast cancer: Long-term risks and predictors,} {\em Journal of
  Cardiac Failure\/}, 20, S44--S45.

\bibitem[\protect\citeauthoryear{Fan, Gao, and Steinhart}{Fan
  et~al.}{2020}]{Fan2020}
Fan, T., Gao, L., and Steinhart, Y. (2020), \enquote{The small predicts large
  effect in crowdfunding,} {\em Journal of Consumer Research\/}, 47, 544--565.

\bibitem[\protect\citeauthoryear{Fine, Jiang, and Chappell}{Fine
  et~al.}{2001}]{Fine2001}
Fine, J.~P., Jiang, H., and Chappell, R. (2001), \enquote{On semi-competing
  risks data,} {\em Biometrika\/}, 88, 907--919.

\bibitem[\protect\citeauthoryear{Gao, Wang, Li, and Cotte}{Gao
  et~al.}{2025}]{Gao2025}
Gao, H., Wang, X.~S., Li, X., and Cotte, J. (2025), \enquote{Crowdfunding
  success for female versus male entrepreneurs depends on whether a consumer
  versus investor decision frame is salient,} {\em Journal of Marketing
  Research\/}, 0, 1--20.

\bibitem[\protect\citeauthoryear{Herd, Mallapragada, and Narayan}{Herd
  et~al.}{2022}]{Herd2022}
Herd, K.~B., Mallapragada, G., and Narayan, V. (2022), \enquote{Do backer
  affiliations help or hurt crowdfunding success?} {\em Journal of
  Marketing\/}, 86, 117--134.

\bibitem[\protect\citeauthoryear{Hofert and Pham}{Hofert and
  Pham}{2013}]{Hofert2013}
Hofert, M. and Pham, D. (2013), \enquote{{Densities of nested Archimedean
  copulas},} {\em Journal of Multivariate Analysis\/}, 118, 37--52.

\bibitem[\protect\citeauthoryear{Hsieh, Wang, and Ding}{Hsieh
  et~al.}{2008}]{hsieh2008regression}
Hsieh, J.-J., Wang, W., and Ding, A.~A. (2008), \enquote{Regression analysis
  based on semicompeting risks data,} {\em Journal of the Royal Statistical
  Society Series B: Statistical Methodology\/}, 70, 3--20.

\bibitem[\protect\citeauthoryear{Hsu and Taylor}{Hsu and
  Taylor}{2009}]{hsu2009nonparametric}
Hsu, C.-H. and Taylor, J.~M. (2009), \enquote{Nonparametric comparison of two
  survival functions with dependent censoring via nonparametric multiple
  imputation,} {\em Statistics in Medicine\/}, 28, 462--475.

\bibitem[\protect\citeauthoryear{Hsu, Taylor, Murray, and Commenges}{Hsu
  et~al.}{2006}]{hsu2006survival}
Hsu, C.-H., Taylor, J.~M., Murray, S., and Commenges, D. (2006),
  \enquote{Survival analysis using auxiliary variables via non-parametric
  multiple imputation,} {\em Statistics in Medicine\/}, 25, 3503--3517.

\bibitem[\protect\citeauthoryear{Huang, Wang, and Emura}{Huang
  et~al.}{2021}]{Huang2021}
Huang, X.-W., Wang, W., and Emura, T. (2021), \enquote{A copula-based Markov
  chain model for serially dependent event times with a dependent terminal
  event,} {\em Japanese Journal of Statistics and Data Science\/}, 4, 917--951.

\bibitem[\protect\citeauthoryear{Jabr and Rahman}{Jabr and
  Rahman}{2022}]{Jabr2022}
Jabr, W. and Rahman, M.~S. (2022), \enquote{Online reviews and information
  overload: The role of selective, parsimonious, and concordant top reviews,}
  {\em MIS Quarterly\/}, 46, 1517--1550.

\bibitem[\protect\citeauthoryear{Joe}{Joe}{2014}]{Joe2014}
Joe, H. (2014), {\em Dependence Modeling with Copulas\/}, Chapman \& Hall/CRC
  Monographs on Statistics and Applied Probability, CRC Press.

\bibitem[\protect\citeauthoryear{Kim, Kannan, Trusov, and Ordanini}{Kim
  et~al.}{2020}]{Kim2020}
Kim, C., Kannan, P.~K., Trusov, M., and Ordanini, A. (2020), \enquote{Modeling
  dynamics in crowdfunding,} {\em Marketing Science\/}, 39, 339--365.

\bibitem[\protect\citeauthoryear{Kozinets, Valck, Wojnicki, and
  Wilner}{Kozinets et~al.}{2010}]{Kozinets2010}
Kozinets, R.~V., Valck, K.~D., Wojnicki, A.~C., and Wilner, S.~J. (2010),
  \enquote{Networked narratives: Understanding word-of-mouth marketing in
  online communities,} {\em Journal of Marketing\/}, 74, 71--89.

\bibitem[\protect\citeauthoryear{Li, Hu, McBride, and Spinelli}{Li
  et~al.}{2020{\natexlab{a}}}]{Li2020}
Li, D., Hu, X.~J., McBride, M.~L., and Spinelli, J.~J. (2020{\natexlab{a}}),
  \enquote{{Multiple event times in the presence of informative censoring:
  Modeling and analysis by copulas},} {\em Lifetime Data Analysis\/}, 26,
  573--602.

\bibitem[\protect\citeauthoryear{Li, Hu, and Wang}{Li et~al.}{2023}]{Li2023}
Li, D., Hu, X.~J., and Wang, R. (2023), \enquote{Evaluating association between
  two event times with observations subject to informative censoring,} {\em
  Journal of the American Statistical Association\/}, 118, 1282--1294.

\bibitem[\protect\citeauthoryear{Li, Chinchilli, and Wang}{Li
  et~al.}{2019}]{Li2019Bayesian}
Li, Z., Chinchilli, V.~M., and Wang, M. (2019), \enquote{A Bayesian joint model
  of recurrent events and a terminal event,} {\em Biometrical Journal\/}, 61,
  187--202.

\bibitem[\protect\citeauthoryear{Li, Chinchilli, and Wang}{Li
  et~al.}{2020{\natexlab{b}}}]{Li2020Bayesian}
--- (2020{\natexlab{b}}), \enquote{A time-varying Bayesian joint hierarchical
  copula model for analysing recurrent events and a terminal event: An
  application to the cardiovascular health study,} {\em Journal of the Royal
  Statistical Society Series C: Applied Statistics\/}, 69, 151--166.

\bibitem[\protect\citeauthoryear{Liang, Li, Li, Chinchilli, Zhang, and
  Wang}{Liang et~al.}{2023}]{Liang2023}
Liang, M., Li, Z., Li, L., Chinchilli, V.~M., Zhang, L., and Wang, M. (2023),
  \enquote{Tackling dynamic prediction of death in patients with recurrent
  cardiovascular events,} {\em Statistics in Medicine\/}, 42, 3487--3507.

\bibitem[\protect\citeauthoryear{Liu, Gao, and Rao}{Liu et~al.}{2025}]{Liu2025}
Liu, Z., Gao, Q., and Rao, R.~S. (2025), \enquote{Self-donations and charitable
  contributions in online crowdfunding: An empirical analysis,} {\em Journal of
  Marketing\/}, 89, 117--134.

\bibitem[\protect\citeauthoryear{Nagler}{Nagler}{2025}]{Nagler2025}
Nagler, T. (2025), \enquote{Simplified vine copula models: State of science and
  affairs,} {\em Risk Sciences\/}, 1, 100022.

\bibitem[\protect\citeauthoryear{Nelsen}{Nelsen}{2006}]{Nelsen2006}
Nelsen, R.~B. (2006), {\em An Introduction to Copulas\/}, New York, NY:
  Springer, 2 edition.

\bibitem[\protect\citeauthoryear{Palmatier, Jarvis, Bechkoff, and
  Kardes}{Palmatier et~al.}{2009}]{Palmatier2009}
Palmatier, R.~W., Jarvis, C.~B., Bechkoff, J.~R., and Kardes, F.~R. (2009),
  \enquote{The role of customer gratitude in relationship marketing,} {\em
  Journal of Marketing\/}, 73, 1--18.

\bibitem[\protect\citeauthoryear{Pan and Joe}{Pan and Joe}{2022}]{Pan2022}
Pan, S. and Joe, H. (2022), \enquote{Predicting times to event based on vine
  copula models,} {\em Computational Statistics \& Data Analysis\/}, 175,
  107546.

\bibitem[\protect\citeauthoryear{Peng and Fine}{Peng and
  Fine}{2007}]{peng2007regression}
Peng, L. and Fine, J.~P. (2007), \enquote{Regression modeling of semicompeting
  risks data,} {\em Biometrics\/}, 63, 96--108.

\bibitem[\protect\citeauthoryear{Sklar}{Sklar}{1959}]{Sklar1959}
Sklar, A. (1959), \enquote{Fonctions de r{\'e}partition {\`a} n dimensions et
  leurs marges,} {\em Publications de l'Institut Statistique de
  l'Universit{\'e} de Paris\/}, 8, 229--231.

\bibitem[\protect\citeauthoryear{St{\"o}ber and Schepsmeier}{St{\"o}ber and
  Schepsmeier}{2013}]{stober2013}
St{\"o}ber, J. and Schepsmeier, U. (2013), \enquote{Estimating standard errors
  in regular vine copula models,} {\em Computational Statistics\/}, 28,
  2679--2707.

\bibitem[\protect\citeauthoryear{Su, Sengupta, Li, and Chen}{Su
  et~al.}{2023}]{Su2023}
Su, L., Sengupta, J., Li, Y., and Chen, F. (2023), \enquote{{``Want'' versus
  ``Need'': How linguistic framing influences responses to crowdfunding
  appeals},} {\em Journal of Consumer Research\/}, 50, 923--944.

\bibitem[\protect\citeauthoryear{Sun and Cook}{Sun and Cook}{2024}]{Sun2024}
Sun, J. and Cook, T. (2024), \enquote{A simple and robust parametric shared
  frailty model for recurrent events with the competing risk of death: An
  application to the Carvedilol Prospective Randomized Cumulative Survival
  trial,} {\em Statistical Methods in Medical Research\/}, 33, 765--793.

\bibitem[\protect\citeauthoryear{Taylor, Murray, and Hsu}{Taylor
  et~al.}{2002}]{taylor2002Survival}
Taylor, J. M.~G., Murray, S., and Hsu, C.-H. (2002), \enquote{Survival
  estimation and testing via multiple imputation,} {\em Statistics \&
  Probability Letters\/}, 58, 221--232.

\bibitem[\protect\citeauthoryear{Tsiatis}{Tsiatis}{1981}]{Tsiatis1981}
Tsiatis, A.~A. (1981), \enquote{A Large Sample Study of Cox's Regression
  Model,} {\em The Annals of Statistics\/}, 9, 93--108.

\bibitem[\protect\citeauthoryear{van~der Vaart and Wellner}{van~der Vaart and
  Wellner}{1996}]{vanderVaart1996}
van~der Vaart, A.~W. and Wellner, J.~A. (1996), {\em Weak Convergence and
  Empirical Processes: With Applications to Statistics\/}, New York, NY:
  Springer.

\bibitem[\protect\citeauthoryear{Vatter and Nagler}{Vatter and
  Nagler}{2018}]{Vatter2018}
Vatter, T. and Nagler, T. (2018), \enquote{Generalized additive models for
  pair-copula constructions,} {\em Journal of Computational and Graphical
  Statistics\/}, 27, 715--727.

\bibitem[\protect\citeauthoryear{Wei, Hong, and Tellis}{Wei
  et~al.}{2022}]{Wei2022}
Wei, Y.~M., Hong, J., and Tellis, G.~J. (2022), \enquote{Machine learning for
  creativity: Using similarity networks to design better crowdfunding
  projects,} {\em Journal of Marketing\/}, 86, 87--104.

\bibitem[\protect\citeauthoryear{Zeng and Lin}{Zeng and Lin}{2006}]{Zeng2006}
Zeng, D. and Lin, D.~Y. (2006), \enquote{{Efficient estimation of
  semiparametric transformation models for counting processes},} {\em
  Biometrika\/}, 93, 627--640.

\bibitem[\protect\citeauthoryear{Zhang and Liu}{Zhang and
  Liu}{2012}]{Zhang2012}
Zhang, J. and Liu, P. (2012), \enquote{Rational herding in microloan markets,}
  {\em Management Science\/}, 58, 892--912.

\bibitem[\protect\citeauthoryear{Zhang, Hua, and Huang}{Zhang
  et~al.}{2010}]{Zhang2010}
Zhang, Y., Hua, L., and Huang, J. (2010), \enquote{{A spline-based
  semiparametric maximum likelihood estimation method for the Cox model with
  interval-censored data},} {\em Scandinavian Journal of Statistics\/}, 37,
  338--354.

\bibitem[\protect\citeauthoryear{Zhao, Wu, and Yin}{Zhao
  et~al.}{2017}]{Zhao2017}
Zhao, X., Wu, Y., and Yin, G. (2017), \enquote{{Sieve maximum likelihood
  estimation for a general class of accelerated hazards models with bundled
  parameters},} {\em Bernoulli\/}, 23, 3385--3411.

\bibitem[\protect\citeauthoryear{Zhu, Lan, Ning, and Shen}{Zhu
  et~al.}{2021}]{zhu2021semiparametric}
Zhu, H., Lan, Y., Ning, J., and Shen, Y. (2021), \enquote{Semiparametric
  copula-based regression modeling of semi-competing risks data,} {\em
  Communications in Statistics-Theory and Methods\/}, 51, 7830--7845.

\end{thebibliography}

\end{document}